\documentclass[fleqn,usenatbib]{mnras}
\usepackage{epsfig,amsmath}
\usepackage{color,subfigure}

\usepackage{mathrsfs,amssymb,amstext,amssymb}
\usepackage{ulem}
\usepackage{url}
\usepackage{bm}
\usepackage{adjustbox}

\usepackage{newtxtext,newtxmath}
\usepackage[T1]{fontenc}
\usepackage{ae,aecompl}

\usepackage{graphicx}	

\def\mean#1{\left< #1 \right>}

\newcommand{\Mpc}{\ensuremath{\,{\rm Mpc}}}

\title[HI bias in the EoR]{The HI Bias during the Epoch of Reionization}

\author[Wenxiao Xu et al.]{
Wenxiao Xu,$^{1,2}$
Yidong Xu$^{1}$,
Bin Yue$^{1}$,
Ilian T Iliev$^{3}$,
Hy Trac$^4$, 
Liang Gao$^{1}$,
Xuelei Chen$^{1,2}$\thanks{E-mail: xuelei@cosmology.bao.ac.cn}
\\
$^{1}$Key Laboratory for Computational Astrophysics, National Astronomical Observatories, Chinese Academy of Sciences, Beijing, 100101, China\\
$^{2}$ School of Astronomy and Space Science, University of Chinese Academy of Sciences, Beijing, 100049, China\\
$^{3}$ Astronomy Centre, Department of Physics \& Astronomy, University of Sussex, Falmer, Brighton, BN1 9QH, UK. \\
$^{4}$ McWilliams Center for Cosmology, Department of Physics, Carnegie Mellon University Pittsburgh, PA 15213, USA
}

\begin{document}
\date{Accepted XXX. Received YYY; in original form ZZZ}
\maketitle

\begin{abstract}
The neutral hydrogen (HI) and its 21 cm line are promising probes to the 
reionization process of the intergalactic medium (IGM). To use this probe effectively, it is imperative to have a good 
understanding on how the neutral hydrogen traces the underlying matter distribution. Here we study this problem using 
semi-numerical modeling by combining the HI in the IGM and the HI from halos during 
the epoch of reionization (EoR), and investigate the evolution and the scale-dependence of the neutral fraction 
bias as well as the 21 cm line bias. We find that the neutral fraction bias on large scales is negative 
during reionization, and its absolute value on large scales increases during the early stage of reionization and then 
decreases during the late stage. During the late stage of reionization, there is a transition scale  at which the HI bias transits
from negative on large scales to positive on small scales, and this scale increases as the reionization
proceeds to the end. 
\end{abstract}



\section{Introduction}

The distribution of neutral hydrogen (HI) in the Universe contains a wealth of cosmological information.
During the dark ages, the HI fluctuations follow the dark matter  density perturbations;
while during the epoch of reionization (EoR), 
the HI is anti-correlated with ionizing sources,
and deviates from the total matter distribution.
After reionization, however, most HI resides in halos, and once again can be used as 
a tracer of the large scale structure (LSS) of matter distribution over a wide range of redshifts and 
scales \citep{Wang2019}. Indeed, the 21 cm intensity mapping technique has been developed to more efficiently map the 
LSS of the Universe using HI as the tracer \citep{Peterson2009}.
This technique is a promising probe of cosmological models (\citealt{Bull2018,Xu2015A,Xu2016A,Obuljen2018b}).

The 21 cm signals from the neutral component of the IGM are almost the unique feasible probe to major epochs of reionization by using current or 
upcoming radio telescopes \citep{Alvarez:2019pss,Furlanetto:2019jzo,Liu:2019srd}, such as the PAPER \citep{Parsons2010}, MWA 
(\citealt{Bowman2013,Tingay2013}), LOFAR \citep{vanHaarlem2013}, HERA \citep{DeBoer2017}, SKA \citep{Koopmans2015}.
However, reionization is a complicated process and at present our understanding is still mostly speculative. 
As other tracers of large scale structure, it is generally believed that on large scales the density of the 
neutral hydrogen should be proportional to the dark matter (DM) density with a nearly constant coefficient which we shall call ``bias''. 
In order to interpret the upcoming 21 cm data from the EoR experiments, it is imperative to know the evolution and 
the scale-dependence of the bias of HI distribution, as well as the physical basis of its behavior.

There have been a number of investigations for the HI bias in the local and post-reionization Universe,  
using both observations and theory. For the low-redshift Universe, 
\citet{Martin2012} analyzed the bias of the HI-selected 
galaxies with $z \lesssim 0.06$ in the ALFALFA survey, and found that on scales 
$\lesssim 10$ $h^{-1}$Mpc, the HI-selected galaxies are anti-biased with respect to DM, while on 
scales $\gtrsim 10$  $h^{-1}$Mpc they are roughly unbiased. This is consistent with an earlier 
analysis for HIPASS samples \citep{Basilakos2007}. \citet{Marn2010} developed an 
analytic framework for the large-scale HI bias based on dark matter halo 
bias and a HI mass-halo mass relation. In \citet{Bagla2010} and \citet{Sarkar2016},  the HI power spectrum 
and the related HI bias are derived by populating HI in halos identified in DM-only simulations. 
The HI bias in a wide redshift range ($0<z<5$) was investigated more explicitly by hydrodynamic 
simulations in \citet{Navarro2018}.  To model the multi-phase hydrogen, they also post-processed 
the star-forming gas. They found that even at $z\sim3$, the HI bias is 
already non-linear at $k\gtrsim 0.3~h$Mpc$^{-1}$. 

During the EoR, however, most of the HI resides in the IGM. 
Various processes such as the formation of early galaxies and the non-linear bubble growth
significantly complicate the bias of the HI distribution from the DM distribution.
HI is not a very good tracer of the DM during this epoch, but it encodes rich 
information of reionization process and its driving sources. \citet{Hoffmann2018} have 
developed a quadratic model for the HI bias, which was found 
to work well during the early stage of reionization, and suggested that measurements of the 
three-point correlation functions in observations can constrain the astrophysical processes 
driving the reionization. \citet{McQuinn2018} developed an effective perturbation theory, 
in which the 21 cm bias is expanded as the sum of terms whose 
coefficients reflect the bias of the ionizing sources, the global neutral fraction, the 
characteristic size of ionized regions and the patchiness of reionization. Still, the HI bias 
during the EoR is much less studied, especially for the transition stage: 
from the late EoR to the post-reionization. During this stage, the HI distribution is 
dominated by both the remaining neutral gas in halos, and the relic voids yet to be ionized, 
i.e., the ``neutral islands'' \citep{Xu2014,Xu2017}. The HI bias may show complicated and 
interesting behaviors. Knowing the HI bias would be convenient  for deriving HI distribution 
statistics from the more simple dark matter distribution in theoretical works. In this work we shall study 
the HI bias during the EoR, especially in the transition era.

This paper is organized as follows. In Section ~\ref{sec:model}, we present a model 
for the HI distribution during the EoR, including both the HI in the IGM from a set of semi-numerical 
simulations, and the HI in halos based on an empirical model.
The results of the evolution and the scale dependence of the HI bias and 21cm bias are presented in 
Section ~\ref{sec:results}. We summarize and discuss our results in Section ~\ref{sec:conclusions}.
Throughout this paper, we assume the $\Lambda$CDM model and adopt the 
following cosmological parameters : 
$\Omega_{\rm b} = 0.045$, $\Omega_{\rm m} = 0.27$, $\Omega_{\Lambda} = 0.73$, 
$h = 0.7$, $\sigma_8 = 0.8$, and $n_{\rm s} = 0.96$.

\section{The HI Model}
\label{sec:model}

During and before the EoR,  the neutral hydrogen exists in both IGM (as large diffuse patches of gas) and halos (as small dense clumps). 
We shall include both components in our model to properly account for the HI distribution.

\subsection{The HI in the IGM}

\begin{figure*}
\centering
\subfigure[z=9.0,\ $\bar{x}^{\rm IGM}_{\rm HI}$=0.82]{
\begin{minipage}[t]{0.5\textwidth}
\centering
\includegraphics[width=3.8in]{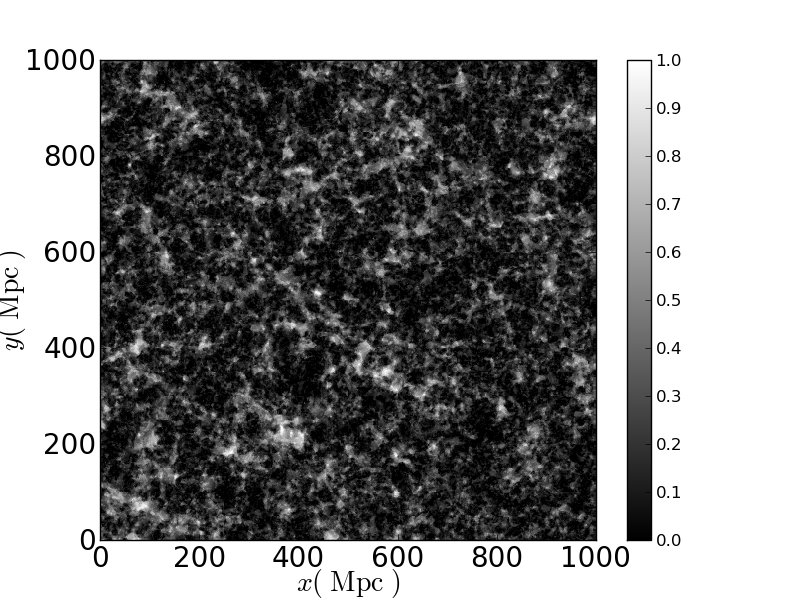}
\end{minipage}%
}%
\subfigure[z=7.8,\ $\bar{x}^{\rm IGM}_{\rm HI}$=0.53]{
\begin{minipage}[t]{0.5\textwidth}
\centering
\includegraphics[width=3.8in]{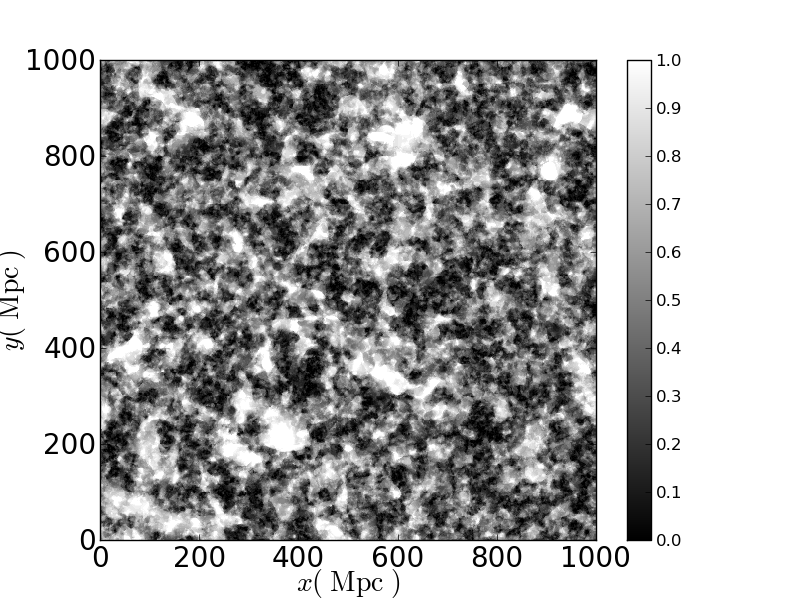}
\end{minipage}%
}\\
\subfigure[z=6.7, \ $\bar{x}^{\rm IGM}_{\rm HI}$=0.14]{
\begin{minipage}[t]{0.5\textwidth}
\centering
\includegraphics[width=3.8in]{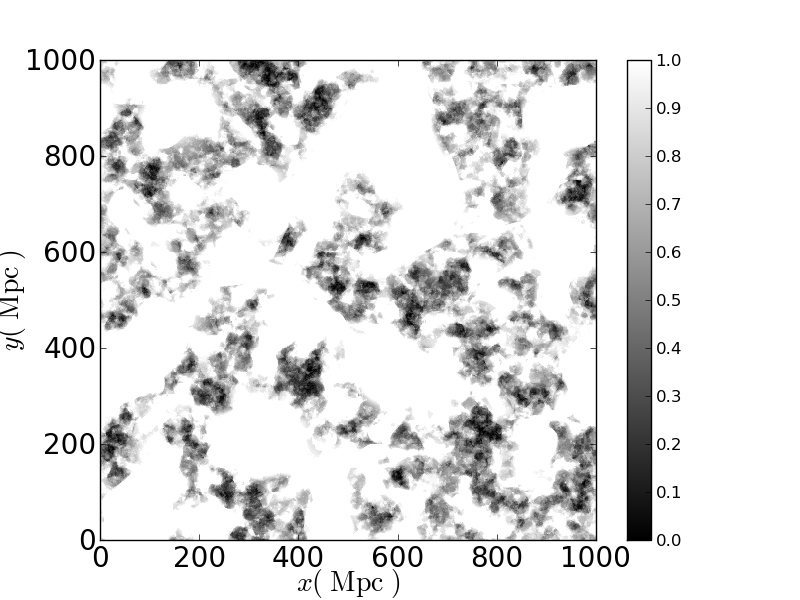}
\end{minipage}%
}%
\subfigure[z=6.5,\ $\bar{x}^{\rm IGM}_{\rm HI}$=0.025]{
\begin{minipage}[t]{0.5\textwidth}
\centering
\includegraphics[width=3.8in]{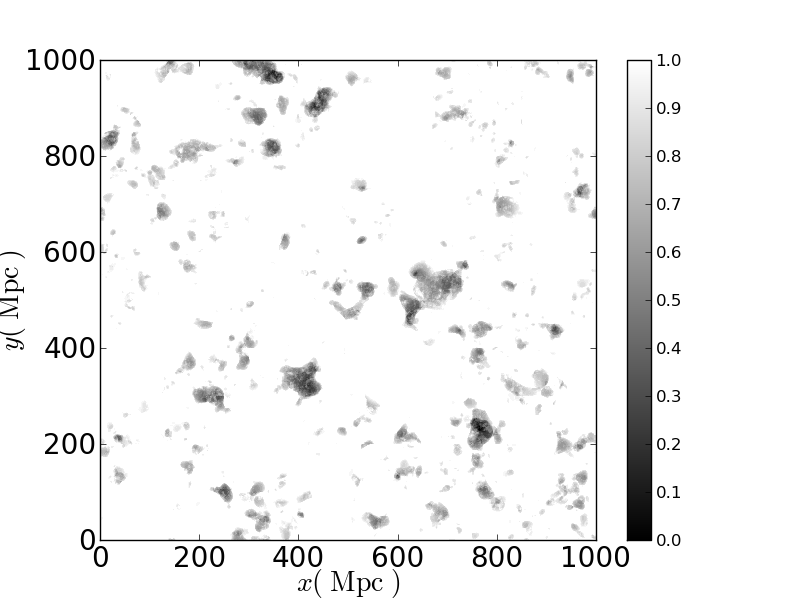}
\end{minipage}
}%
\centering
\caption{A visualization of the evolution of the ionized field of the simulation (1Gpc). The mean neutral fractions for the four panels are
 0.82, 0.53, 0.14 and 0.025 respectively. 
}
\label{fig:field}
\end{figure*}

It is generally believed that reionization started first in high density regions where the first luminous objects formed first. 
In the ``bubble model''  of reionization \citep{Furlanetto2004}, the amount of star formation and the resulting ionizing photons are 
estimated using the excursion set model. Based on this idea, the so called ``semi-numerical simulations" have been developed. 
For example,  \citet{Mesinger2011} developed the {\tt 21cmFAST}\footnote{https://github.com/andreimesinger/21cmFAST}, to
simulate the evolution of the 3D density, ionization and 21cm brightness temperature fields efficiently.  
The ``bubble model'' considers spherical regions of increasingly smaller scales
and identifies ionized bubbles by comparing the cumulative number of ionizing photons 
produced within the region with the number consumed in reionization process. 
It has been demonstrated that the statistical predictions of the ``bubble model'' and the {\tt 21cmFAST} 
agree fairly well with radiative-hydrodynamic simulations \citep{Zahn2007,Mesinger2011,Zahn2011}, at least when recombination, 
feedback, etc. are ignored. However, during most epochs of reionization, the topology of the ionization field is much more complicated than 
the isolated bubbles configuration. The bubbles start to connect with each other as early as 
when the global ionized fraction is just about 10\%, and the Universe starts the percolation
process when the ionized fraction gets $\sim$ 30\% (see e.g. \citealt{Furlanetto2016,Chen:2018enj}).
Inspired by the bubble model, the {\tt 21cmFAST} determines the ionization state of each point by comparing the expected 
ionizing photon production in the surrounding region with the required number, but it allows non-spherical geometry for the ionized regions.
In order to give a better description of the evolution of neutral regions after percolation, 
\citet{Xu2014} developed the so-called ``island model'', assuming isolated  neutral islands topology. 
The island model also takes into account an ionizing background that is inevitable during the late EoR
\citep{Furlanetto2005,Emberson2013,McQuinn2011}. 
Based on the ``island model'', a semi-numerical code named {\tt islandFAST} was developed to mimic 
the islands evolution during the last stage of reionization \citep{Xu2017}. 
In the {\tt islandFAST}, the effect of small-scale absorbers is taken into account empirically by adopting
a fitting formula for the evolution of mean free path (MFP) of ionizing photons \citep{2010ApJ...721.1448S}, based on the 
observed number density of Lyman limit systems up to redshift 6. Before the completion of reionization, the MFP is limited
by both the under-dense islands and the over-dense absorbers. The evolution of the ionization field and the intensity of the
ionizing background are derived self-consistently by an iterative procedure to ensure convergence in the total effective
MFP of the ionizing photons.

Topological analysis shows that the islands become mostly isolated when the global neutral 
fraction reduces to $\bar{x}_{\rm HI}\sim 0.16$ in a model with parameters similar to the present case \citep{Chen:2018enj}. 
In the island model the early stage of reionization is identical to that of
the bubble model ({\tt 21cmFAST}) until $\bar{x}_{\rm HI} \approx 0.17$, slightly before the isolation of individual islands, after 
which the island model criterion is applied for the later stages of reionization.
For the following analysis, our simulation has a box size of 1 Gpc (comoving) on a side and 
a resolution of $500^3$ cells. An ionizing efficiency parameter of $\zeta = 20$ and a minimum 
virial temperature of $10^4$ K are adopted for host halos of ionizing sources.
A few slices of the ionization field at several snapshots are shown in Fig.~\ref{fig:field} .

\subsection{The HI in halos}
\label{sec:massfunction}
To model the HI in halos, for each cell in the simulation box with a size of $V$ and a mass of $M$ at redshift $z_0$, 
the number density of halos with mass $m$ virialized at $z_{1}$ in this cell, $n(m,z_{1}|M,V,z_{0})$, 
can be estimated by using the conditional mass function \citep{Cooray2002}:
\begin{eqnarray}
    \frac{m^{2}n(m,z_{1}|M,V,z_{0})}{\bar{\rho}_m}\frac{dm}{m} &=& \nu_{10}f(\nu_{10})\frac{d\nu_{10}}{\nu_{10}}, 
	\label{eq:conditional}
\end{eqnarray}
where 
\begin{eqnarray}
    \nu_{10} &=& \frac{[\delta_{\rm sc}(z_{1})-\delta_{0}(\delta, z_{0})]^{2}}{\sigma^{2}(m)-\sigma^{2}(M)}.
	\label{eq:nu}
\end{eqnarray}
Here $\bar{\rho}_m=\rho_c\Omega_m$ is the mean matter density of the present Universe and $\rho_c=3H^2_0/8\pi G$ is the critical density; $\delta_{\rm sc}(z_{1})$ is critical density required for spherical collapse at $z_{1}$, 
$\delta_{0}(\delta, z_{0})$ denotes the initial density for a region to have density $\delta$ at $z_{0}$,
$\sigma^{2}(m)$ is the variance of the density fluctuations on scale $m$,
and we adopt the form of $f(\nu)$ given in \citet{Sheth1999}.
The relation between the HI mass and its host halo mass has been widely 
studied \citep{Gong2011, Popping2015,Guo2017,Padmanabhan2017,Navarro2018,Obuljen2018a}. 
Here we adopt the results from a large state-of-the-art hydrodynamic simulation,
 TNG100 \citep{Navarro2018}, for which the result can be fitted by
\begin{eqnarray}
    m_{\rm HI}(m, z) &=& m_{0}\left(\frac{m}{m_{\rm min}}\right)^{\alpha}\exp\left({-\frac{m_{\rm min}}{m}}\right).
	\label{eq:himass}
\end{eqnarray}
The parameters are given for a number of redshifts. At $z\sim5$, $m_{\rm 0} = 9.5 \times 10^{7} h^{\rm -1}M_{\odot}$, 
$m_{\rm min} = 1.9 \times 10^{9} h^{\rm -1}M_{\odot}$, and $\alpha = 0.9$ \citep{Navarro2018}. 
We adopt these $z\sim5$ parameter values throughout the reionization history, as that is the highest redshift where the 
values are given, and also the relations vary little near $z\sim5$. Another possibility is to fit the redshift dependence 
at $z\lesssim5$, then extrapolate to higher redshifts. We have checked that such extrapolation gives similar results, 
but given the uncertainty,  it is not clear if a better accuracy can be achieved. 
The HI content of a halo may also depends on its assembly history and when the local patch was ionized.
Obviously, there are always some theoretical 
uncertainties in the model of HI gas in halos during the EoR, 
both intrinsic and environmental, which one must bear in mind.

The neutral fraction of a given cell with mass $M$ and volume $V$ contributed by HI in halos can be written as
\begin{equation}
    x^{\rm halo}_{\rm HI} (M,V,z) = \frac{1}{\rho_c\Omega_{\rm b}X_{\rm H}}\int_0^\infty n(m,z|M,V,z)
    m_{\rm HI}(m, z)\, {\rm d}m,
	\label{eq:rho}
\end{equation}
where $X_{\rm H}$ donates the hydrogen mass fraction in the gas, $X_{\rm H} \sim 0.75$. 
The HI in halos are combined with the HI in the IGM, predicted by semi-numerical simulations such as 
 the {\tt 21cmFAST} or {\tt islandFAST}, to generate 
the full HI field at various redshifts throughout the reionization.
In the following, we denote the neutral fraction in the IGM as $x^{\rm IGM}_{\rm HI}$, 
the neutral fraction contributed from the HI in halos as $x^{\rm halo}_{\rm HI}$, and denote
the total neutral fraction as $f_{\rm HI}$.

\section{The HI Bias throughout the EoR}
\label{sec:results}

The formation of galaxies and the evolution of the IGM are physical 
consequences of the primordial density perturbations. On large scales,  the observables such as 
the galaxy number density, the neutral hydrogen fraction of the IGM or the 21cm brightness temperature should be related to the 
locally averaged matter density. Here we approximate such relation with a linear model, 
 though more generally the relation may be non-linear and stochastic \citep{1999ApJ...520...24D}. 

The reionization process is believed to be ``inside-out'' (e.g. \citealt{Iliev2006,Trac2007}) 
on large scales: galaxies formed firstly in over-dense regions, and ionized their surrounding gas, 
then the reionization proceeds from the over-dense regions to the mean and under-dense regions. 
In this scenario, it is expected that the neutral hydrogen is anti-correlated with the underlying dark matter 
density distribution on large scales. However, this relation could be inverted on small scales, where the 
neutral hydrogen in halos is correlated with the density, and also the ionization front moves faster into the void regions.  
The non-linear structure growth, galaxy formation, ionizations and feedbacks would significantly
complicate the neutral hydrogen-density relation and hence the HI bias. The time- and scale-dependence
of the HI bias encodes rich information about the reionization process.

The neutral hydrogen is observationally traced by the 21 cm signal, the 
upcoming low-frequency interferometers will directly measure the fluctuations in the 21 cm brightness temperature.
Ignoring the peculiar velocities, the 21 cm brightness temperature is related to 
the neutral fraction $f_{\rm HI}$, the density contrast $\delta$, and the spin temperature $T_{\rm s}$
via (e.g. \citealt{Pritchard2012})
\begin{equation}
    \delta T_{\rm b} = 27\, f_{\rm HI}(1+\delta)\left(\frac{\Omega_{\rm b}h^{2}}{0.023}\right)
    \left(\frac{0.15}{\Omega_{\rm m}h^{2}} \frac{1+z}{10}\right)^{1/2}\left(\frac{T_{\rm s} - T_{\gamma}}{T_{\rm s}}\right)\, {\rm mK},
	\label{eq:21cmbrightness}
\end{equation}
where $T_{\gamma}$ is the background photon temperature, and in the absence of very strong radio sources it would be given by the 
cosmic microwave background temperature at that redshift. Although when the first stars formed $T_{\rm s}$ might be lower than 
$T_{\gamma}$ \citep{2004ApJ...602....1C,2008ApJ...684...18C},  after a moderate fraction of gas had been ionized, the rest of the neutral HI would 
be heated to $T_{\rm s} \gg T_{\gamma}$. For simplicity here we shall assume this relation holds through the whole EoR process.
To calculate the contributions to the 21cm signal from halos and the IGM separately, one can replace the $f_{\rm HI}$ with $x_{\rm HI}^{\rm halo}$ 
or with $x_{\rm HI}^{\rm IGM}$. Strictly speaking, Eq. (\ref{eq:21cmbrightness}) is only valid in the optically-thin limit, which may not be true 
for HI gas in halos. However, it has been shown that for the average of $\delta T_b$  from many halos, it is still 
a good enough approximation \citep{Yue2009}.

  The neutral fraction bias can be computed by comparing the power spectra of  
the neutral fraction field and dark matter density field, either from the auto-power spectrum
\begin{eqnarray}
    b^2_{f}(k) &\equiv& P_{\rm HI}(k)/P_{\rm \delta \delta}(k),
	\label{eq:bias1}
\end{eqnarray}
or from the cross-power spectrum
\begin{eqnarray}
    b_{f}(k) &\equiv& P_{\rm HI \delta}(k)/P_{\rm \delta \delta}(k);
	\label{eq:bias2}
\end{eqnarray}
where
\begin{eqnarray}
P_{\rm \delta \delta}(k)&=&\mean{\hat{\delta}_m(k) \hat{\delta}^*_m(k)}^\prime,\nonumber\\
P_{\rm HI}(k)&=&\mean{ \hat{f}_{\rm HI} (k) \hat{f}^*_{\rm HI} (k)}^\prime,\\
P_{\rm HI \delta}(k)&=&\mean{\hat{f}_{\rm HI}(k) \hat{\delta}^*_m(k)}^\prime\nonumber
\end{eqnarray}
 are the matter power spectrum, neural fraction power spectrum, and cross-power spectrum between the two, 
 respectively,  $\hat{f}_{\rm HI}$ and $\hat{\delta}(k)$ are the Fourier transform of $f_{\rm HI}$ and dark matter density contrast respectively, and $\mean{}^\prime$ denotes the  the expectation value with the Dirac delta function removed.
  The bias defined in this way can be negative, this is similar to the bias of voids \citep{Sheth:2003py,Chan:2014qka}.
We shall use Eq. (\ref{eq:bias2}) which is less affected by the stochasticity and the shot noise term \citep{Navarro2018}, and also there is no 
ambiguity in the sign of the bias. 
Similarly, we also compute the 21cm bias using 
\begin{eqnarray}
b_{21}(k)&\equiv& P_{T\delta}(k)/P_{\delta\delta}(k),
\label{eq:bias3}
\end{eqnarray}
where $\hat{\delta}_T(k)$ is the Fourier transform of $(\delta T_b-\mean{\delta T_b})/\mean{\delta T_b}$ and 
$P_{T\delta}(k)=\mean{\hat{\delta}_T(k) \delta^*_m(k)}$. Throughout this paper for the 21cm power spectrum or 21cm-dark matter 
cross-power spectrum we adopt the dimensionless definition.

\subsection{The power spectra}\label{sec:ps}

\begin{figure}
\centering
\includegraphics[width=0.5\textwidth]{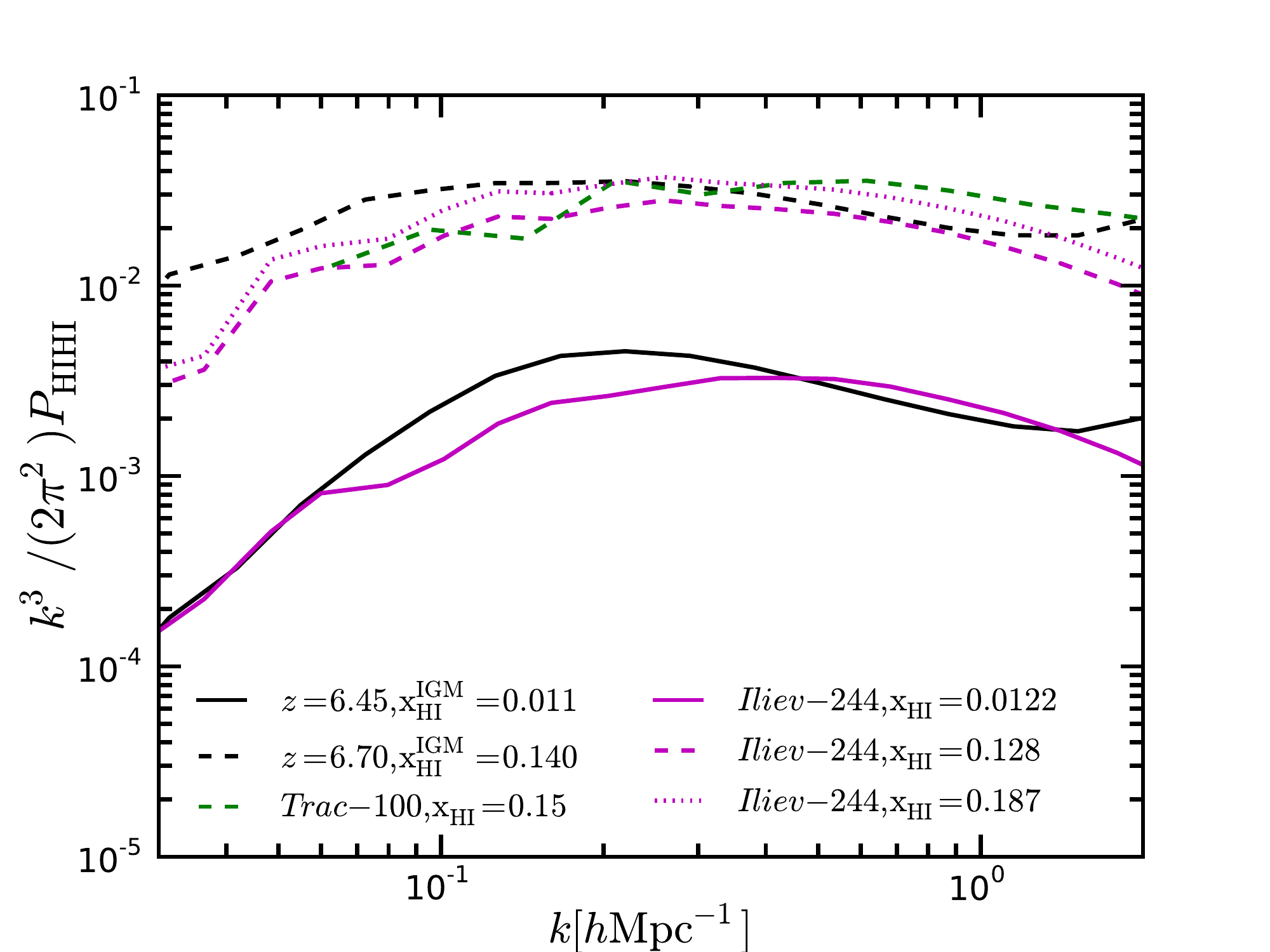}
\includegraphics[width=0.5\textwidth]{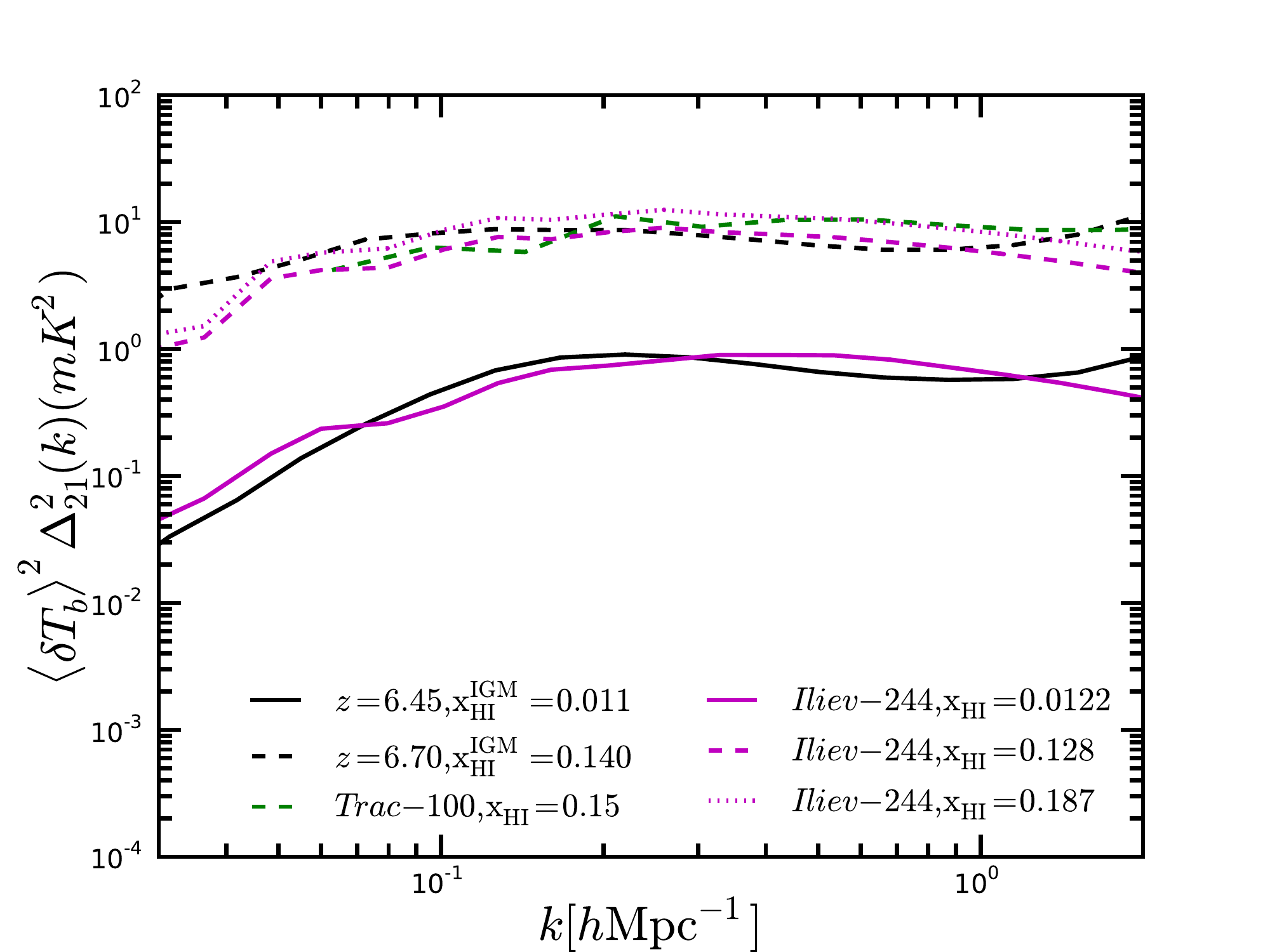}
\caption{The neutral fraction ({\it upper panel}) and the 21 cm 
power spectra ({\it lower panel}) predicted by the {\tt islandFAST} (black lines) as compared with the one
predicted by \citet{BTCL2013} (green line, denoted by ``Trac-100'' in the legend) and 
the ones from \citet{2016MNRAS.456.3011D} (magenta lines, denoted by ``Iliev-244''). 
The mean neutral fraction of each line is indicated in the legend.}
\label{fig:island21cmps}
\end{figure}

\begin{figure}
\centering      
\subfigure[Dark matter power spectrum]{
\begin{minipage}[t]{0.5\textwidth}
\centering
\includegraphics[width=0.9\textwidth]{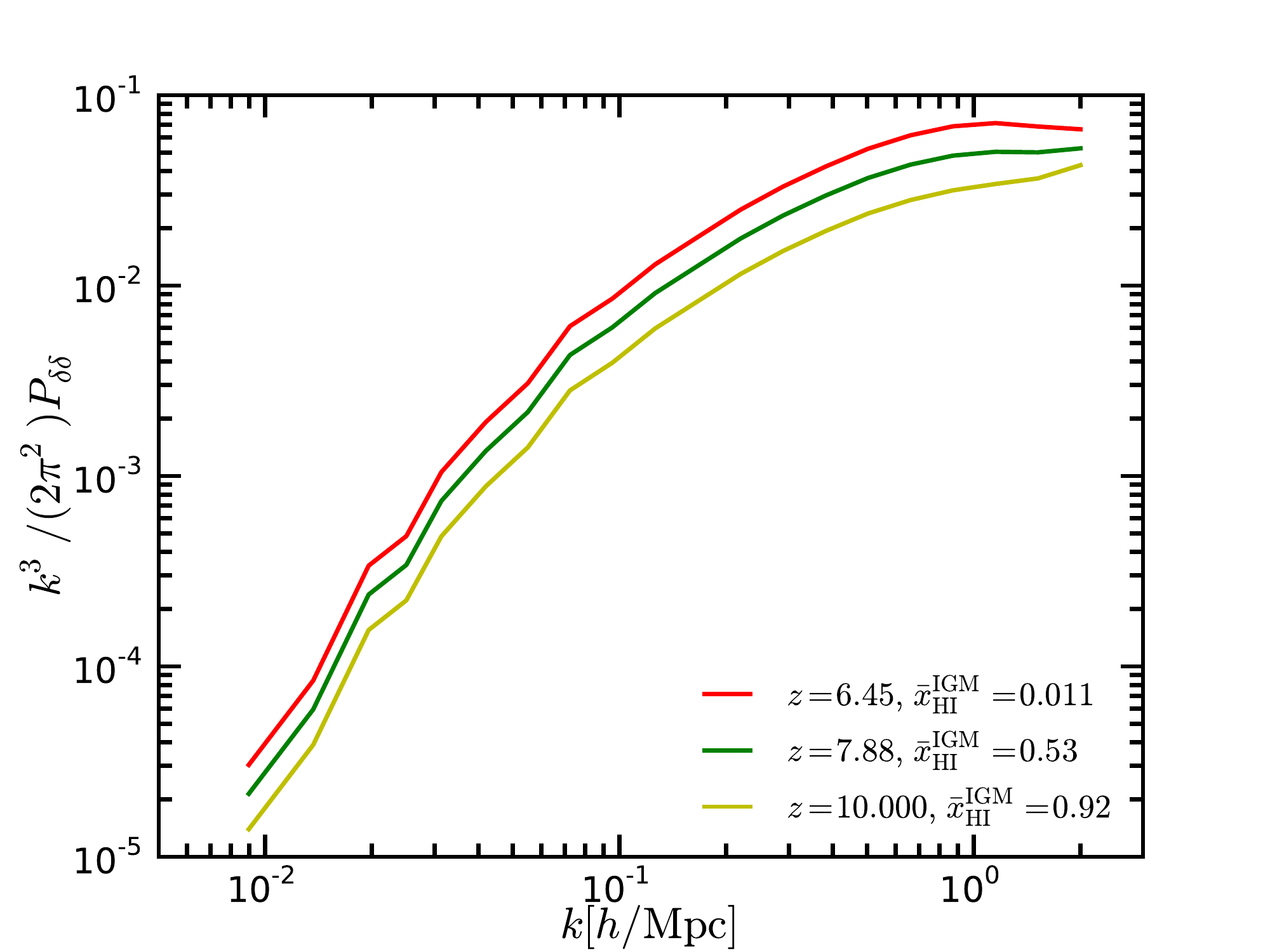}
\end{minipage}%
}
\subfigure[Neutral fraction power spectrum]{
\begin{minipage}[t]{0.5\textwidth}
\centering
\includegraphics[width=0.9\textwidth]{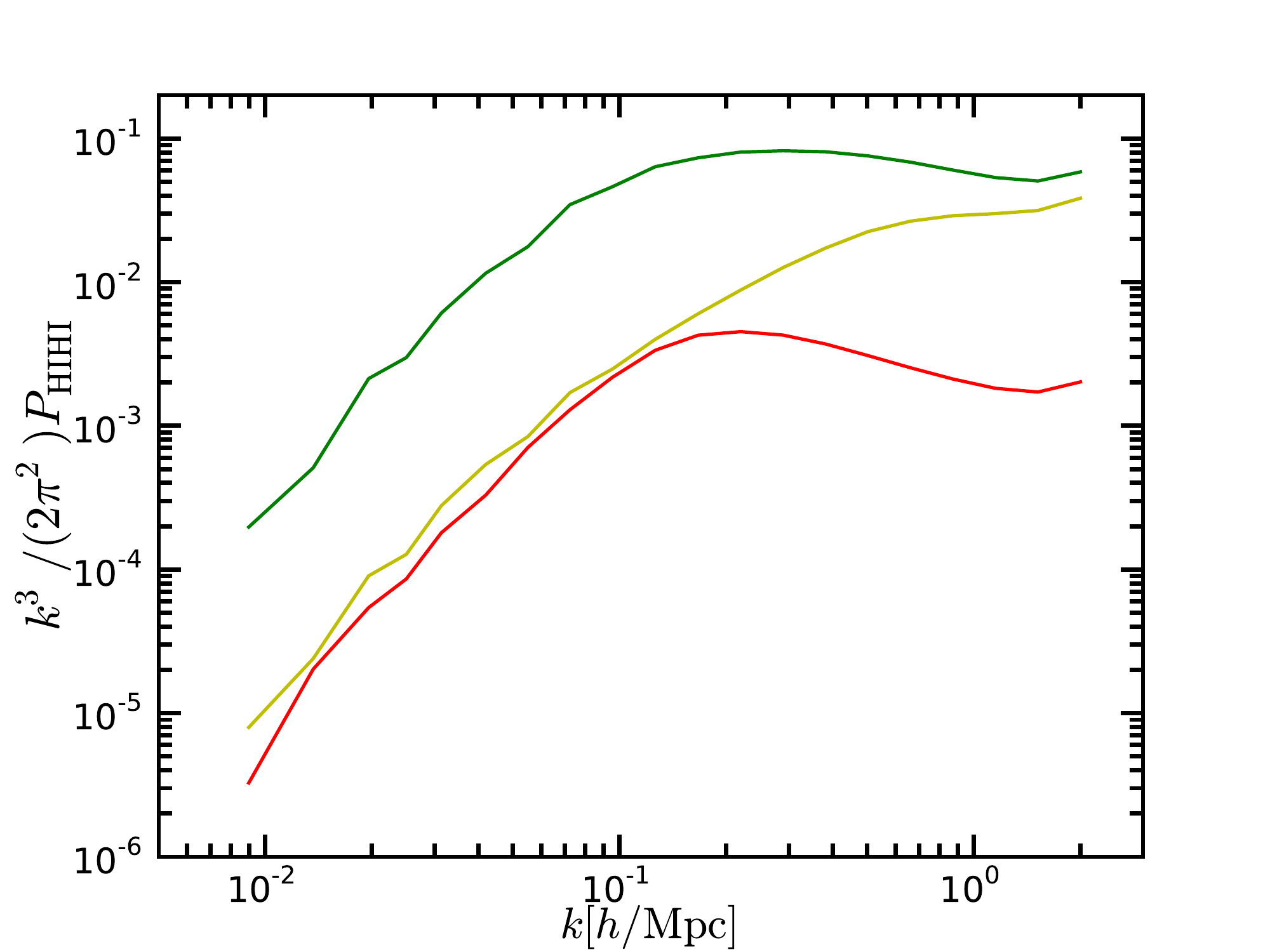}
\end{minipage}%
}
\subfigure[21 cm power spectrum]{
\begin{minipage}[t]{0.5\textwidth}
\centering
\includegraphics[width=0.9\textwidth]{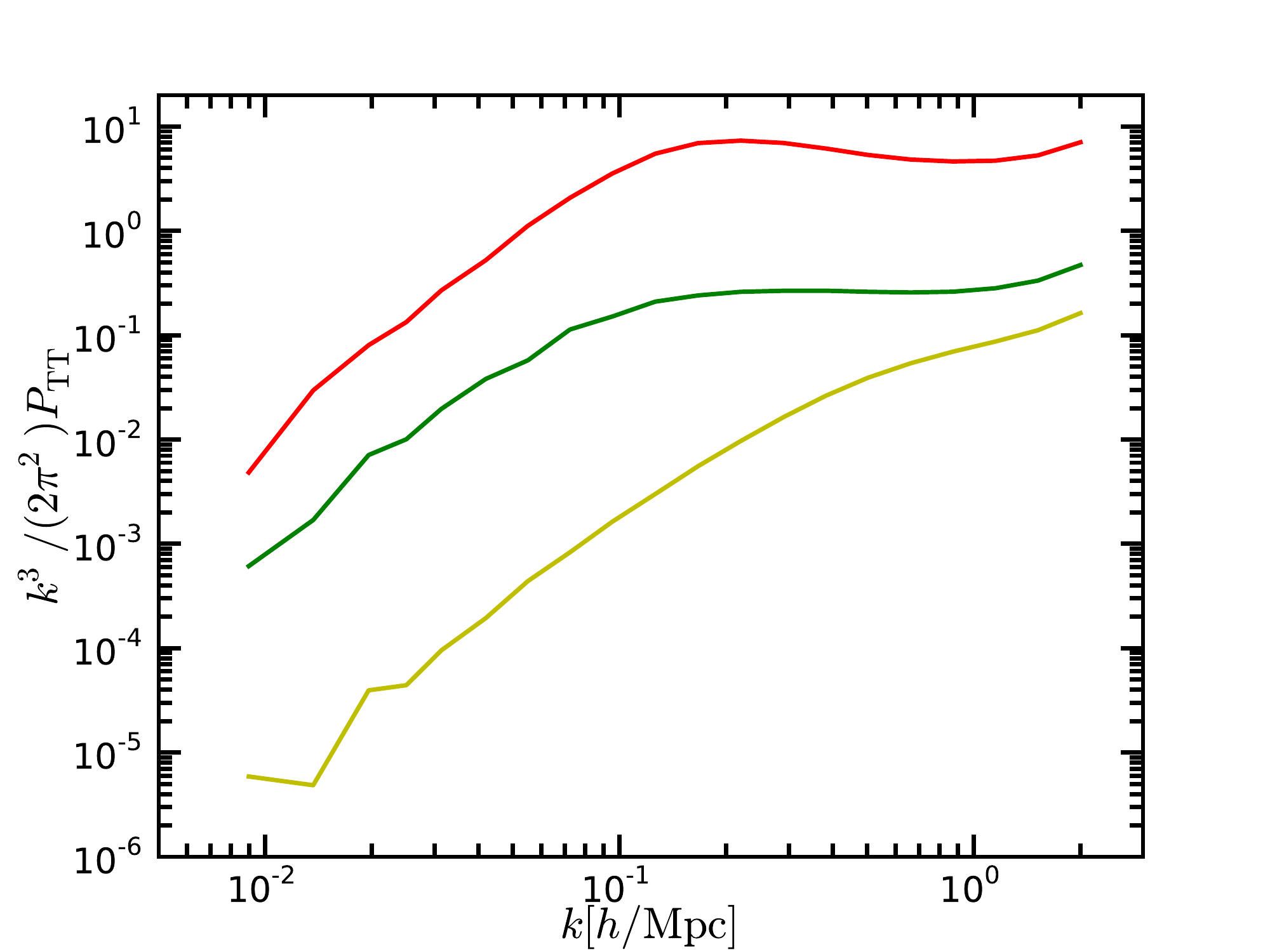}
\end{minipage}%
}
\centering
\caption{The power spectra of dark matter ({\it top panel}), 
the HI neutral fraction $f_{\rm HI}$ ({\it middle panel}), 
and the 21 cm brightness temperature ({\it bottom panel}). 
The three curves are taken at $z = 10.0$, 7.88 and 6.45, corresponding to IGM mean  
neutral fraction $\bar{x}^{\rm IGM}_{\rm HI} = $0.92, 0.53, and 0.011 respectively. 
}
\label{fig:ps1}
\end{figure}

In the present work, the reionization process is simulated by the {\tt 21cmFAST} for early stage and by 
the {\tt islandFAST} after percolation.
The 21 cm power spectrum predicted by the {\tt 21cmFAST} was previously presented and discussed in 
\citet{Mesinger2011}. While the 21 cm power spectrum during the late stages of reionization was also
discussed in \citet{2019arXiv190301294G},
here we compare the {\tt islandFAST} predictions with the radiative-hydrodynamic simulations by 
\citet{BTCL2013} (Trac-100 simulation) and by \citet{2016MNRAS.456.3011D} (Iliev-244 simulation) 
at some similar neutral fractions. The neutral fraction power spectra and the 21cm power spectra,
in the limit of $T_{\rm s} \gg T_{\gamma}$, 
are plotted in  the upper and lower panels of Fig.~\ref{fig:island21cmps}, respectively.
The black solid and dashed lines 
correspond to neutral fractions of $x_{\rm HI} \sim 0.01$ and 0.14, respectively 
from {\tt islandFAST}. 
Here the box sizes are $200~h^{-1} \Mpc$ for the 
Trac-100 simulation (green line), and $244~h^{-1} \rm{Mpc}$ for the Iliev-244 simulation 
(magenta lines), respectively.
Generally, these neutral fraction and the 21 cm power spectra have similar shape on large scales. 
 Considering the differences between the two numerical
 simulations and output neutral fractions, these differences are probably not very significant, though
the inclusion of {\it evolutional} small scale ionizing photons absorbers in the {\tt islandFAST} 
semi-numerical code may also contributed to the difference. 

The dark matter density power spectra (e.g.\citealt{2010MNRAS.406.2421S,2014MNRAS.439..725I}), 
as well as the 21 cm power spectra (e.g. \citealt{Mesinger2007,2016MNRAS.455.4498K}) during the EoR 
have been studied in a number of previous simulations. 
In this subsection, we calculate the power spectra from our simulation, in the same convention 
as the ones in the bias formula (Eq.(\ref{eq:bias2}) -- (\ref{eq:bias3}) ). 
In addition to the matter density and 21 cm power spectra, here we also calculate the 
neutral fraction power spectra, as well as the cross-power between them, 
in order to gain a better understanding of the bias behavior to be presented in the next subsections.

In Fig.~\ref{fig:ps1} we plot the dark matter power spectra (top panel), the power spectra for
the neutral fraction $f_{\rm HI}$ field (middle panel) and 21cm brightness temperature field (bottom 
panel) in our model during the EoR. In the top panel, the three curves from bottom to top are for the early, 
intermediate and late stages of reionization, respectively, and in the middle and bottom panels 
the corresponding curves are plotted with the same colors as the top panel.

 The power spectrum of the dark matter density fluctuations 
steadily increase as the redshift decreases, while keeping the shape almost unchanged on the scales
discussed here.  By contrast, neither the redshift evolution nor the scale dependence of  the 
neutral fraction power spectrum is monotonic. The neutral fraction power spectrum first increases during the 
early stage of reionization, and then decreases during the late stage of reionization. The 
21cm power spectrum, however, keeps increasing with decreasing redshift until  $x_{\rm HI}^{\rm IGM}\sim0.01$.  Note here 
we are referring to the dimensionless power, and part of the increase of the power comes from the decreasing  mean 21cm brightness temperature
as more regions are ionized. In the late stage of reionization, there is a bump around $k\sim 0.1~h$/Mpc 
on both of them. This scale is close to the typical ionized bubble size, and the HI fluctuations at smaller scales
are somewhat suppressed by reionization process. 

\begin{figure}
\centering
\includegraphics[width=0.5\textwidth]{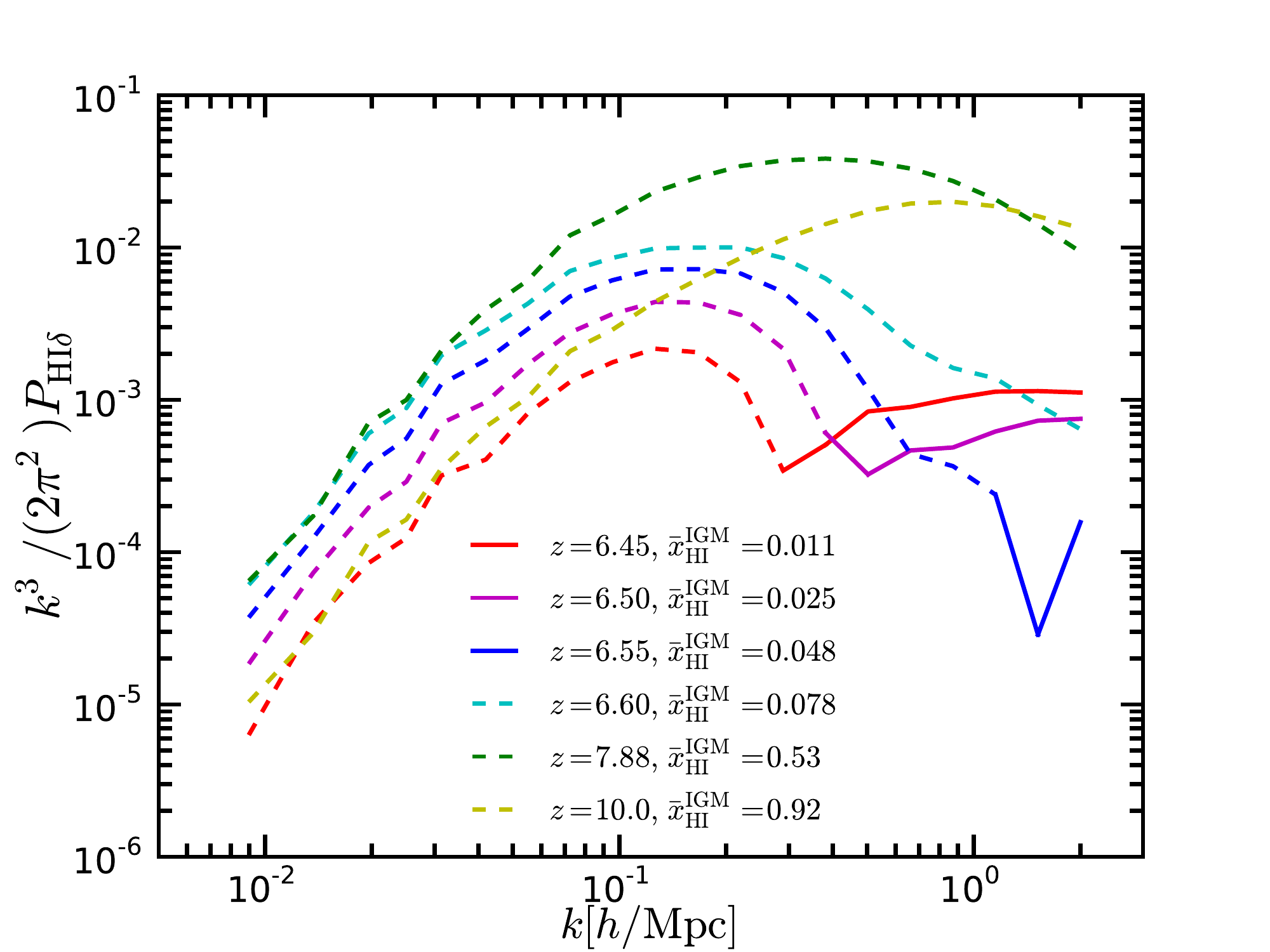}
\caption{The cross-power spectrum between the neutral fraction and the dark matter density. 
The solid and the dashed lines represent positive and negative values respectively. 
}
\label{fig:ps_HIpower}
\end{figure}

The cross-power spectrum between the neutral fraction field and the dark matter density field is shown in 
Fig.\ref{fig:ps_HIpower}.
We use dashed lines to mark the negative power, while solid lines refer to positive power.
As seen in this figure, during most of the EoR stages, the neutral fraction field is anti-correlated
with the dark matter density field, resulting in negative cross-power spectrum. 
As the reionization proceeds, the amplitude of the large scale cross-power spectrum increases during the early stage of
reionization, reaches its maximum roughly at the mid point of reionization, then decreases during the late stage.
When the IGM global neutral fraction drops below $\sim 5 \%$, on the small scales a positive cross-correlation appears. 
At this stage, most of the IGM has been ionized, while the neutral hydrogen in halos starts to dominate the cross power spectrum,
producing the positive correlation seen in the small scale power spectrum.  
As the remaining neutral islands in the IGM shrink and decrease in number, the halo HI  come to dominate the fluctuations on larger and 
larger scales, as a result this positive-to-negative transition point on the cross-power spectrum moves to larger scales. 

\begin{figure}
\centering
\includegraphics[width=0.5\textwidth]{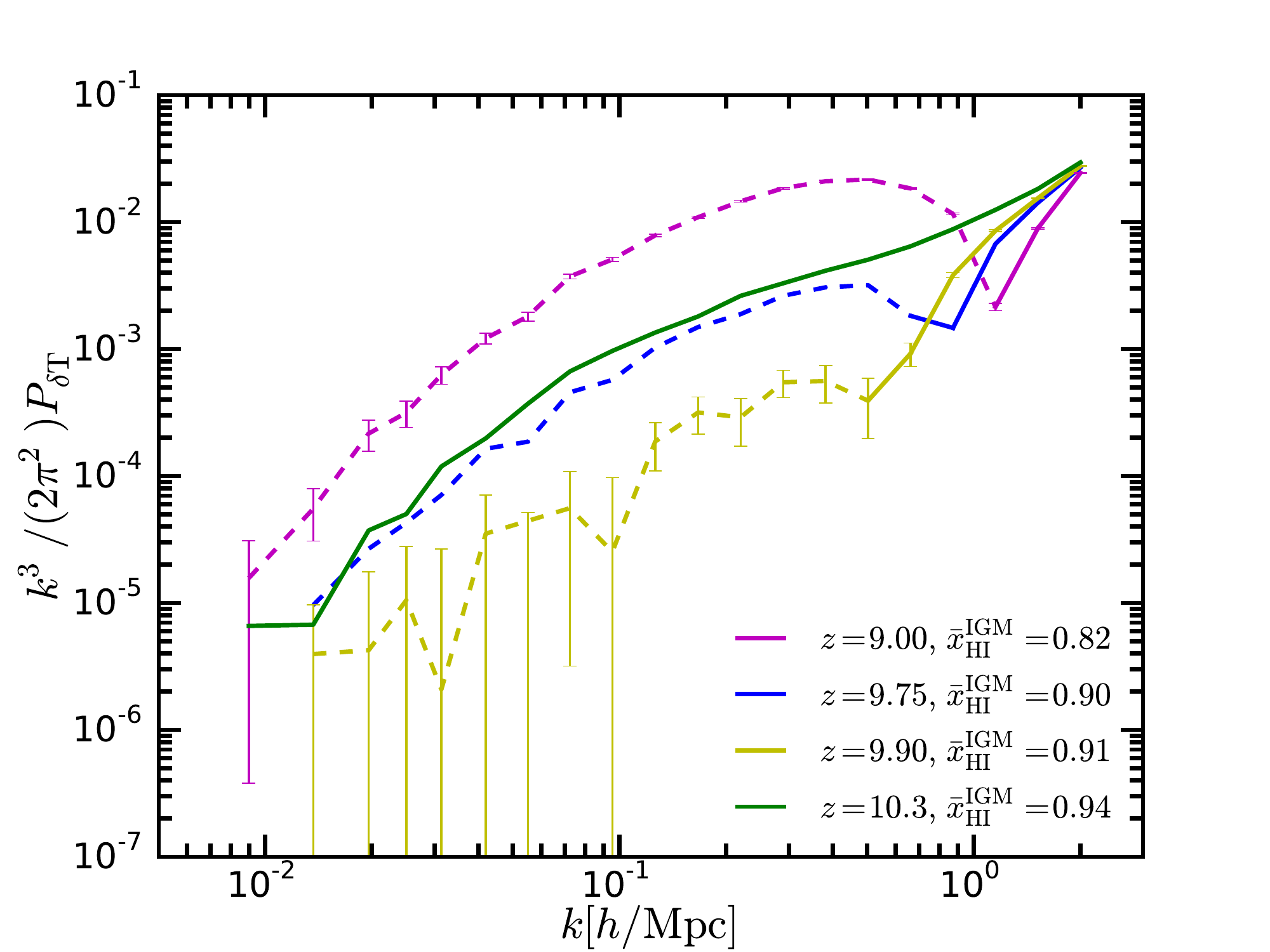}
\includegraphics[width=0.5\textwidth]{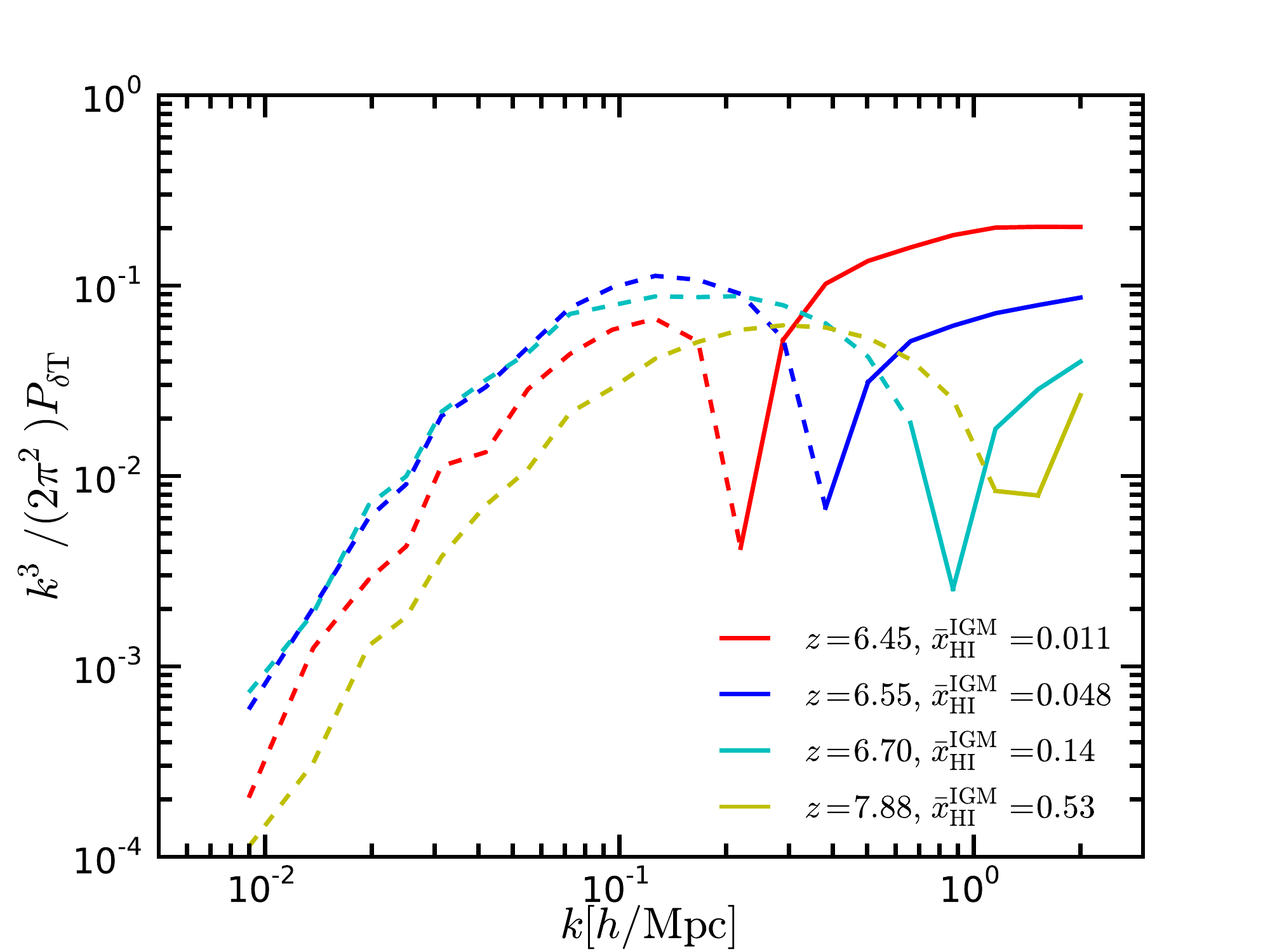}
\caption{The 21 cm brightness temperature and the dark matter density cross power spectrum. 
Top panel:  early EoR; Bottom panel:  late EoR. Solid and dashed lines represent positive and negative values respectively. 
}
\label{fig:ps2}
\end{figure}

\begin{figure}
\centering
\includegraphics[width=0.5\textwidth]{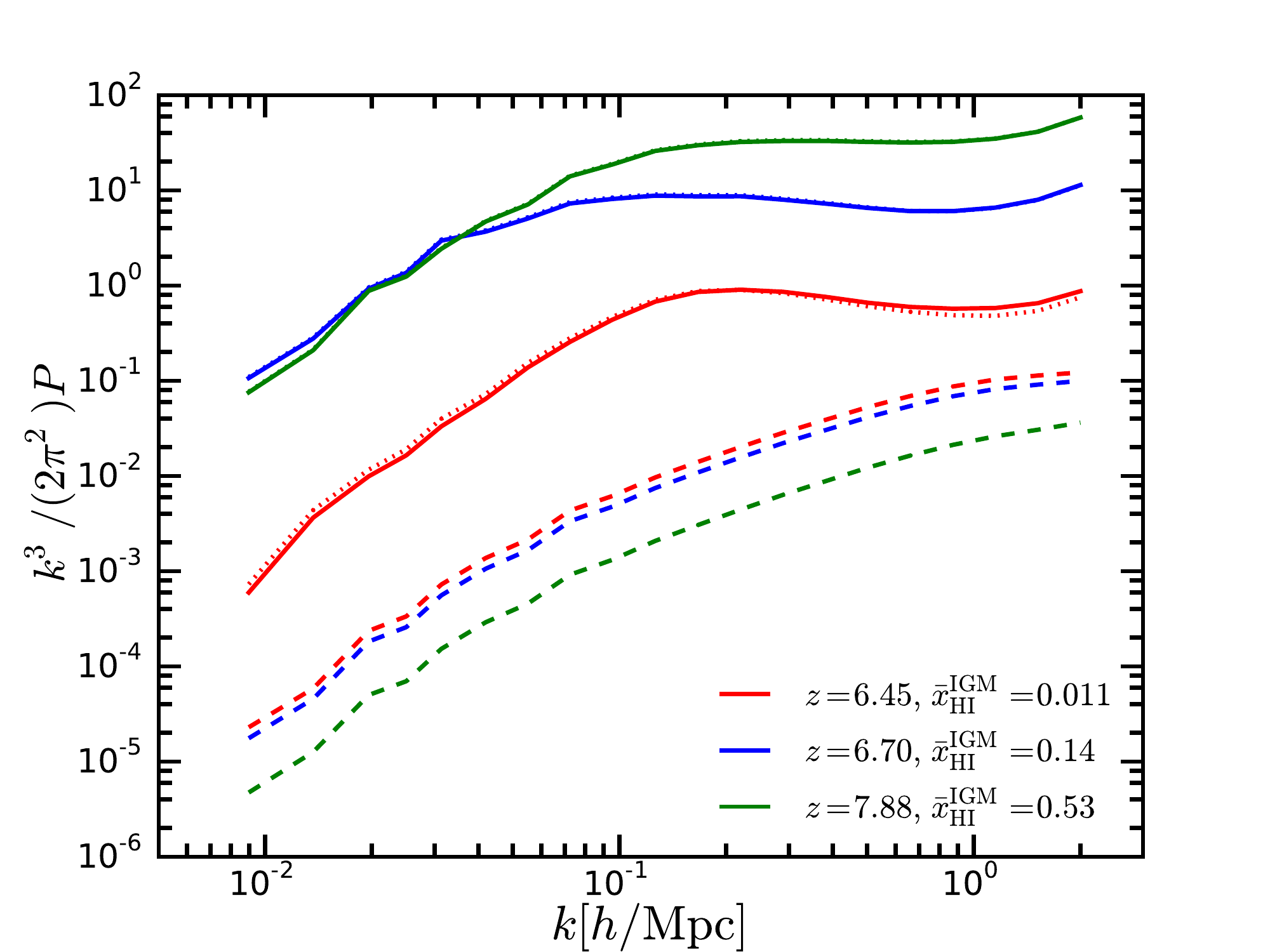}
\centering
\caption{The 21cm power spectrum contributed from the HI in the IGM (dotted lines), the one from the HI in halos (dashed lines), and the total 21cm power spectrum (solid lines).}
\label{fig:antoPS}
\end{figure}

The cross-power spectrum between the 21cm brightness temperature and the dark matter density 
are plotted in Fig.~\ref{fig:ps2}. We use the two panels to show the early and late stages of reionization respectively.
We find that at the very beginning of reionization, the 21cm-dark matter cross-power spectrum is positive on all scales,
implying that the neutral hydrogen density fluctuations still follow the dark matter density fluctuations, although some rare
density peaks have been ionized. As more and more IGM around density peaks becomes ionized, the 21cm-dark matter
cross-power spectrum decreases, and transits to negative starting from the largest scales.
Then the large scale 21cm-dark matter cross-power becomes more and more negative. The amplitude reaches its maximum
at around $x_{\rm HI}^{\rm IGM}\sim0.1$, and decreases again. Different from the $f_{\rm HI}$-dark matter cross-power, the 21cm-dark matter cross-power spectrum is always positive at small scales, indicating that when weighted with the local density, the small scale 21 cm signal is always dominated by the HI in halos.
At the very late stage of reionization, the positive-to-negative transition scale increases rapidly.

In this plot the 21cm-dark matter cross-power spectrum at $z=9.90$ shows some oscillations, which are likely statistical fluctuations. 
To check this interpretation,  we estimate the errors on cross-power spectra by using the jackknife resampling method. 
In Fig. 5 we plot the errors for the z=9.90 and z=9.00 curves. 
 To avoid clutter, errors of other curves are not shown there. We see that indeed at z=9.90 the error bars are larger than the 
 oscillation feature on  the power spectrum curve, so it is likely that these are just statistical fluctuations. We also check that at z=9.00 the errors are 
comparable with the z=9.90. However, at z=9.00 the cross-power spectrum is much larger than z=9.90. In logarithmic axes apparently
 its errors are less obvious than z=9.90.

\begin{figure}
\centering
\includegraphics[width=0.5\textwidth]{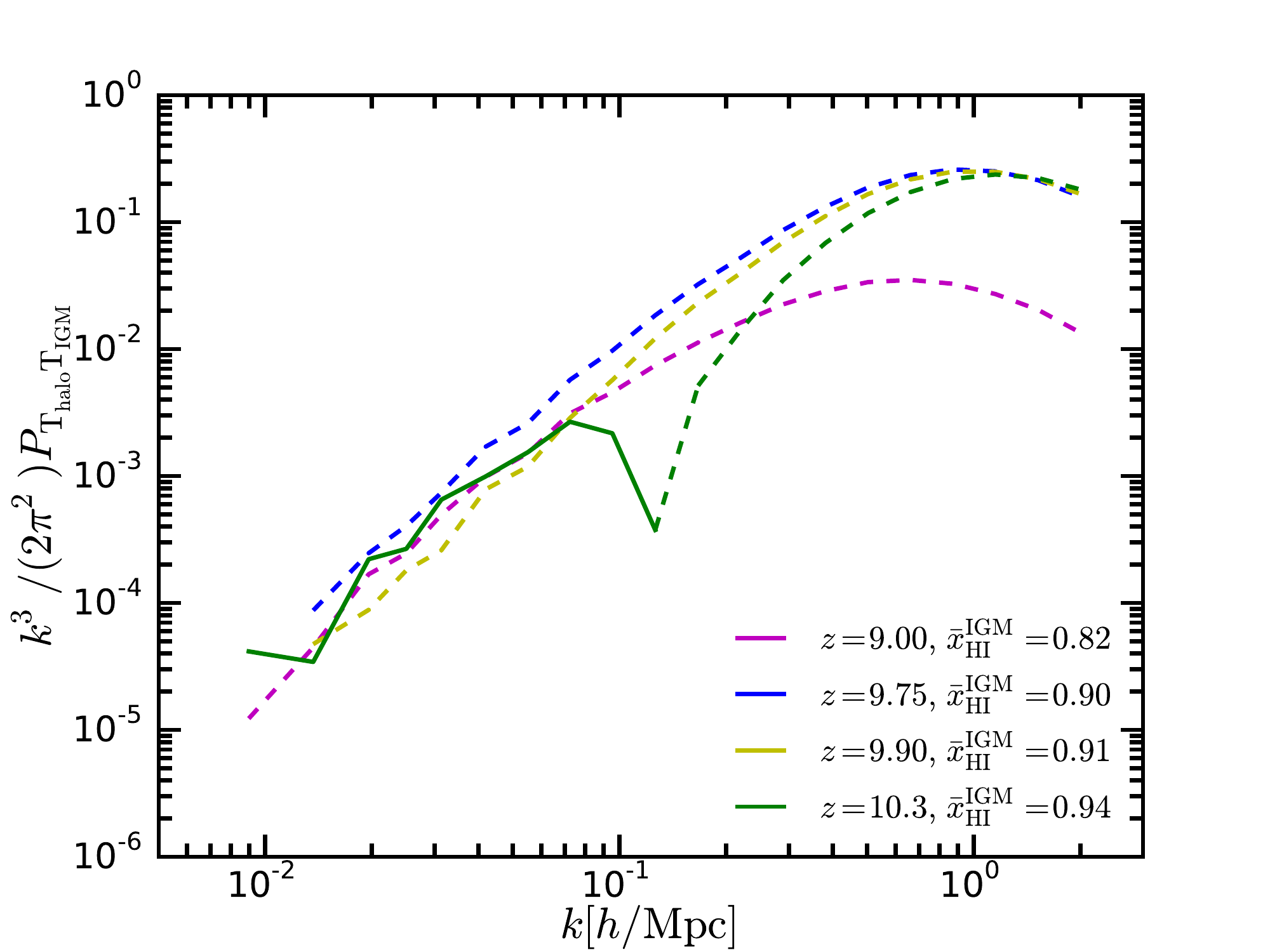}
\includegraphics[width=0.5\textwidth]{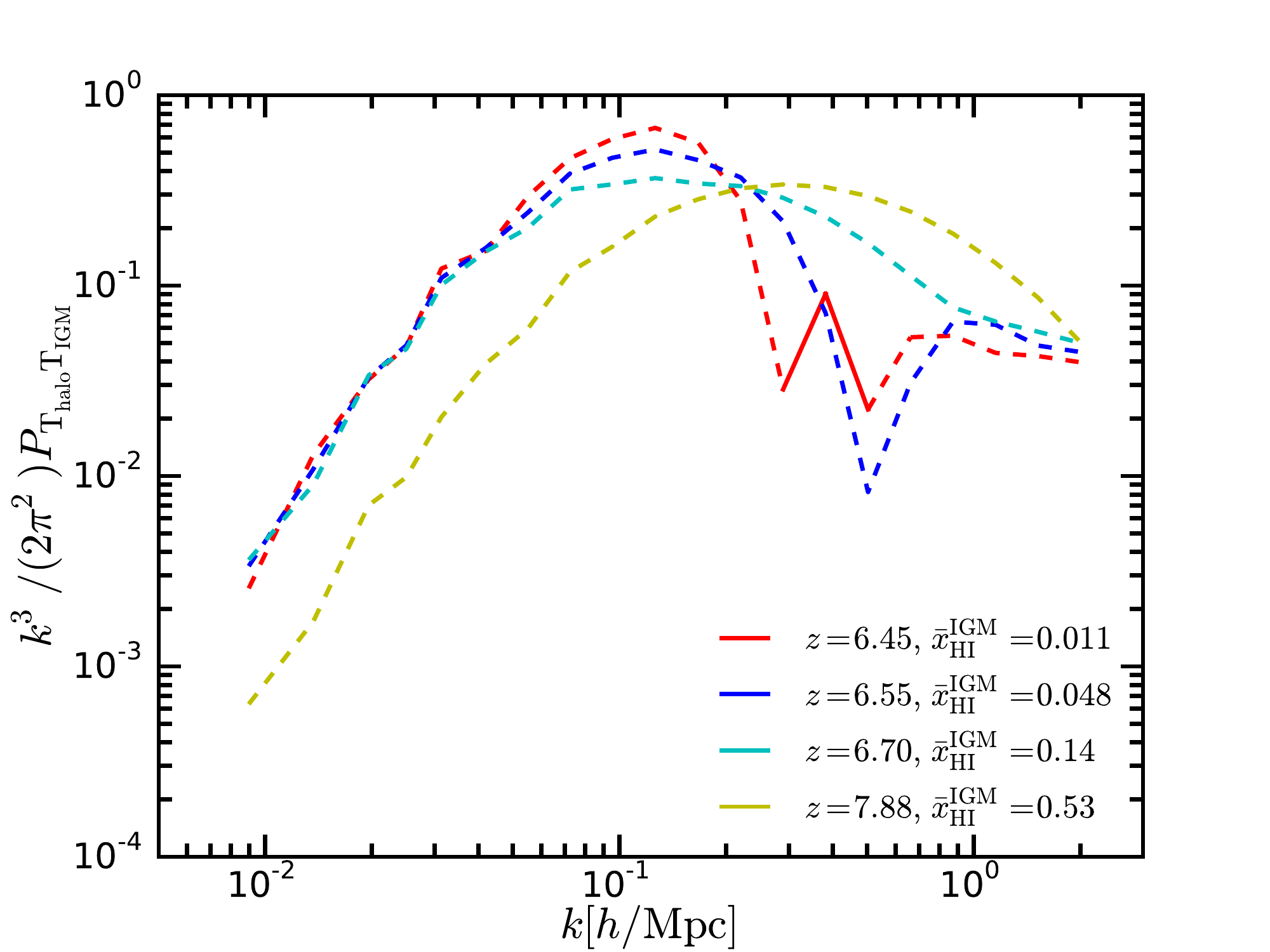}
\caption{The cross-power spectrum of the 21cm brightness from the HI in the IGM and that in halos.
The top panel shows the cross-power spectrum during the early EoR, and the bottom panel shows the
results during the late EoR. The solid and the dashed line represent positive and negative values respectively.  
}
\label{fig:crossPS3}
\end{figure}

We then turn to the individual contribution from the IGM and halos to the 21cm power spectrum, see  Fig.~\ref{fig:antoPS}.
The power spectrum of the HI in the IGM evolves markedly, in contrast to the power spectrum of the HI in halos. 
During the era studied here, the full 21cm power spectrum is dominated by the IGM contribution in our model, though we note here
that we assumed a constant $m_{\rm HI}-m$ relation, if it evolves this conclusion would be revised, though the evolution in this relation 
is perhaps still much less significant than that in the IGM. 

Fig.~\ref{fig:crossPS3} shows the cross-power between the 21cm brightness contributed from the IGM and 
that from halos during the early (top panel) and late EoR (bottom panel). Generally these two components are anti-correlated 
during the EoR, except at the very early stage ($x_{\rm HI}^{\rm IGM} \gtrsim 0.94$) when the ionized bubbles only occupy a tiny 
fraction, there are still positive cross-correlation on large scales ($k \lesssim 0.1$ $h${\rm Mpc}$^{-1}$).

\subsection{The neutral fraction bias $b_{f}$}

\begin{figure}
\centering
\includegraphics[width=0.5\textwidth]{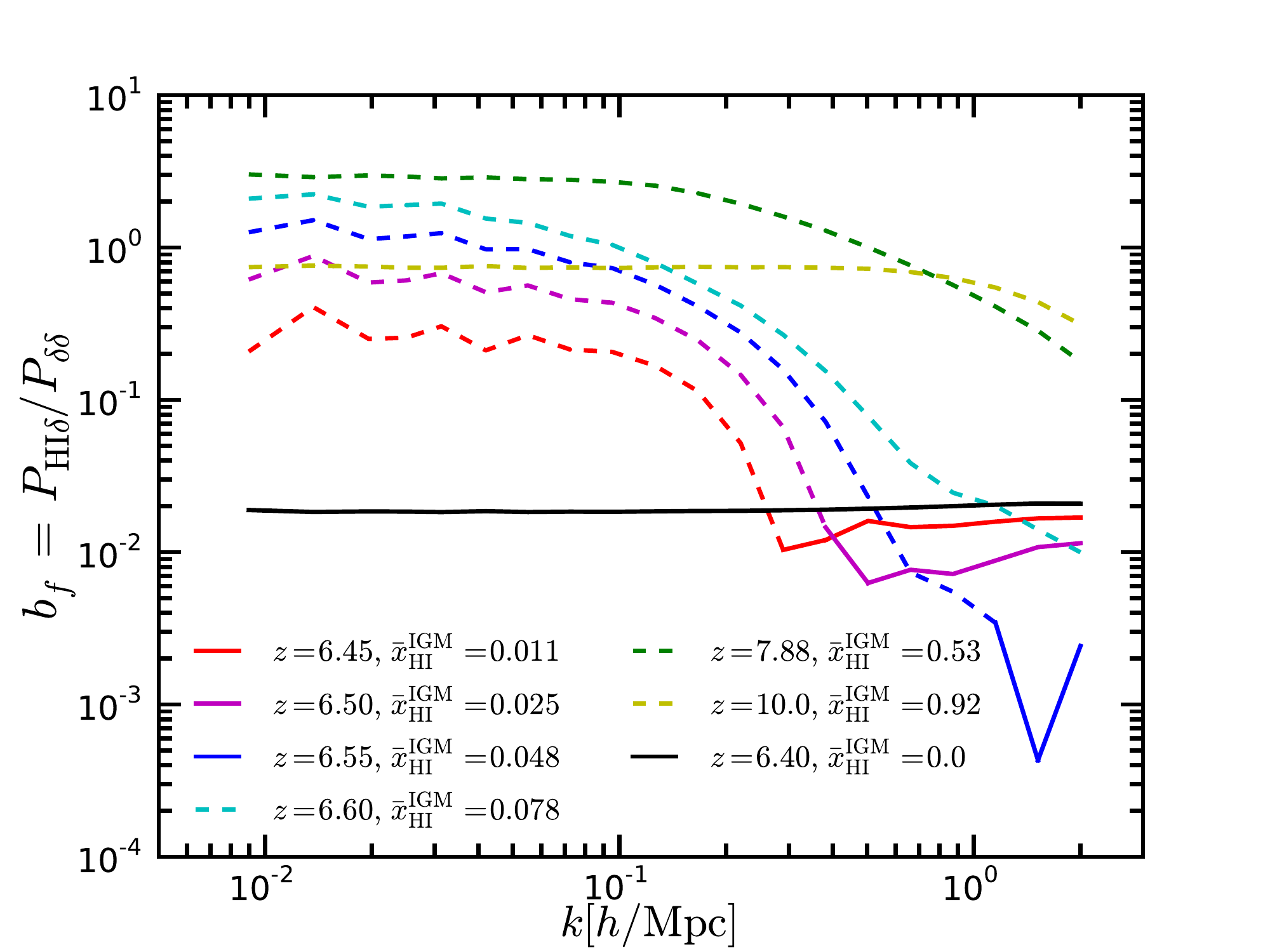}
\caption{The HI bias at various redshifts.
}
\label{fig:ps_HIbias}
\end{figure}

In Fig.~\ref{fig:ps_HIbias}, we show the scale dependence of the neutral fraction bias 
at various stages of reionization. 
As the dark matter power spectrum evolves almost self-similarly throughout the EoR, the evolution of 
neutral fraction bias $b_{f}$ basically follows the trend in the cross-power spectrum $P_{\rm HI \delta}$.
As discussed above, during most of the EoR the HI in the IGM makes a negative contribution to the bias, 
while HI in halos makes a positive contribution. Before the end of reionization, the total bias is negative 
on large scales. When $x_{\rm HI}^{\rm IGM}\gtrsim0.08$, it is negative on all scales, which is 
expected in the ``inside-out'' model of the reionization process. In over-dense regions, 
the HI {\it fraction} is lower than in the under-dense regions, resulting in an anti-correlated 
$f_{\rm HI}-\delta$ relation. Note however that the absolute amount of the remaining HI content 
in the over-dense regions is not necessarily smaller than in the under-dense regions. 
Regarding the $k$-dependence of the $b_f(k)$, there is a typical scale,  below which $b_f$ drops 
gradually, while almost independent of $k$ above this typical scale.  
This typical scale increases as reionization process goes on, reflecting the bubble growth process during the EoR.  

We do expect the bias $b_f$ to be scale-independent on large scales: if we smooth the density field 
by a filter much larger than ionized bubbles, the density field becomes rather smooth, i.e. $\bar{\delta} \ll 1$, where $\bar{\delta}$ is the mean density contrast within the smoothing radius. 
 In such case for a region with mass $M$ and mean density contrast $\bar{\delta}$ the mean neutral fraction $x_{\rm HI}(\bar{\delta})\sim1-\zeta/(1+n_{\rm rec}) f_{\rm coll}(\bar{\delta})$, or $\Delta x_{\rm HI}(\bar{\delta})\propto -\Delta f_{\rm coll}(\bar{\delta})$, where 
\begin{equation}
f_{\rm coll}(\bar{\delta},M,z)={\rm erfc}\left[ \frac{\delta_c(z)-\bar{\delta}}{ \sqrt{2(\sigma^2(M_{\rm min})-\sigma^2(M)) } } \right]
\end{equation}
is the collapse fraction and $n_{\rm rec}$ is the mean recombination number per H atom.
For $\bar{\delta} \ll 1$ we have $\Delta f_{\rm coll} \propto \bar{\delta}$,  and as a result at large scale the HI bias is broadly scale-independent.

When $x_{\rm HI}^{\rm IGM}\gtrsim0.08$, HI in the IGM is dominant and the halo HI has a negligible contribution to the total HI. 
During this stage, on small scales the amplitude of the negative bias decreases rapidly as reionization proceeds. However, on 
large scales, at the beginning the amplitude is smaller, because $f_{\rm HI}$ is rather uniform and close to 
$\sim1$ almost everywhere. As $x_{\rm HI}^{\rm IGM}$ decreases , the $b_f$ amplitude first increases to a peak amplitude $\sim 3$
 when the Universe is about half-ionized, then decreases as well. When $x_{\rm HI}^{\rm IGM}$ decreases to 
 below $\sim0.08$, HI in halos starts to dominate the small scale $f_{\rm HI}$ fluctuations and the positive $b_f$ 
 starts to appear at the small scales. A transition scale, at which the $b_f$ transits from negative at larger scales to positive at 
 smaller scales, increases with the neutral islands being ionized and the increasing contribution by HI in halos.

We also plot the $b_f$ after the completion of reionization in Fig.~\ref{fig:ps_HIbias}.
As expected, after the HI in the IGM gets all ionized, the remaining HI in halos positively correlate
with the underlying matter density, resulting in the positive bias at all scales after reionization.
Note that after reionization, the $b_f$ is basically a weighted halo bias on scales larger than 
the cell size of our simulation. As we have neglected the rare halos with masses larger than the cell mass 
and smoothed the HI density within cells, we have equivalently neglected the one-halo term contribution 
to the power spectrum, and retained only the two-halo term on large scales \citep{Cooray2002}, 
which results in an under-estimate of the bias on small scales.
Also, we note that the halo model, as well as our method of populating simulation cells using the conditional 
mass function, would under-estimate the halo bias on scales  $k\gtrsim 1 \,{\rm h^{-1}\, Mpc}$ 
as compared to that from N-body simulations. Both of these effects give rise to 
the nearly scale-independent bias on the relevant scales.
Interestingly, the neutral fraction bias $b_f$ transits from fully negative to fully positive in a 
very short time scale, which is a clear indication of the completion of reionization.

\begin{figure}
\centering
\subfigure[21 cm bias during early EoR]{
\begin{minipage}[t]{0.5\textwidth}
\centering
\includegraphics[width=1.0\textwidth]{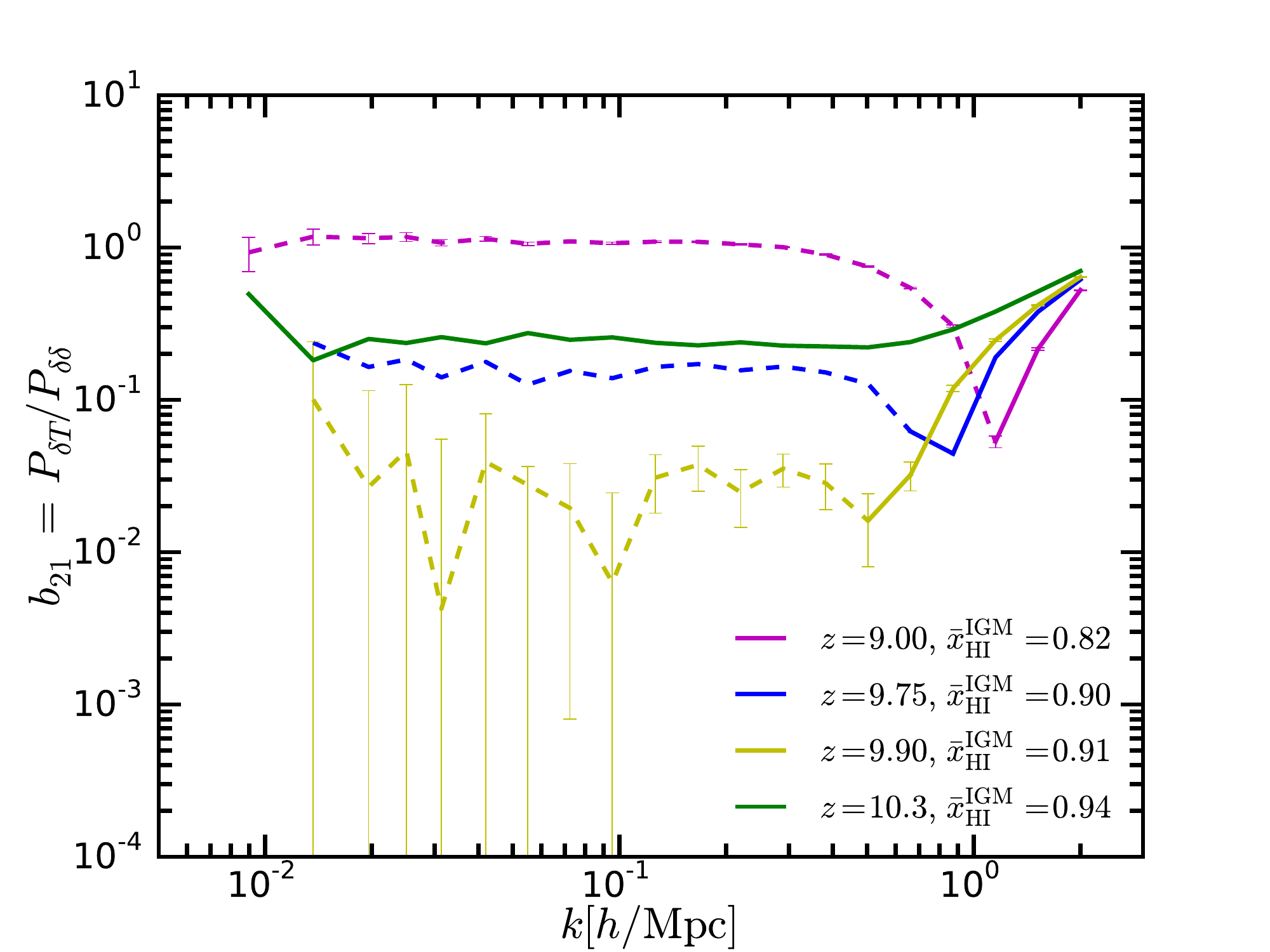}
\end{minipage}%
}\\
\subfigure[21 cm bias during late EoR]{
\begin{minipage}[t]{0.5\textwidth}
\centering
\includegraphics[width=1.0\textwidth]{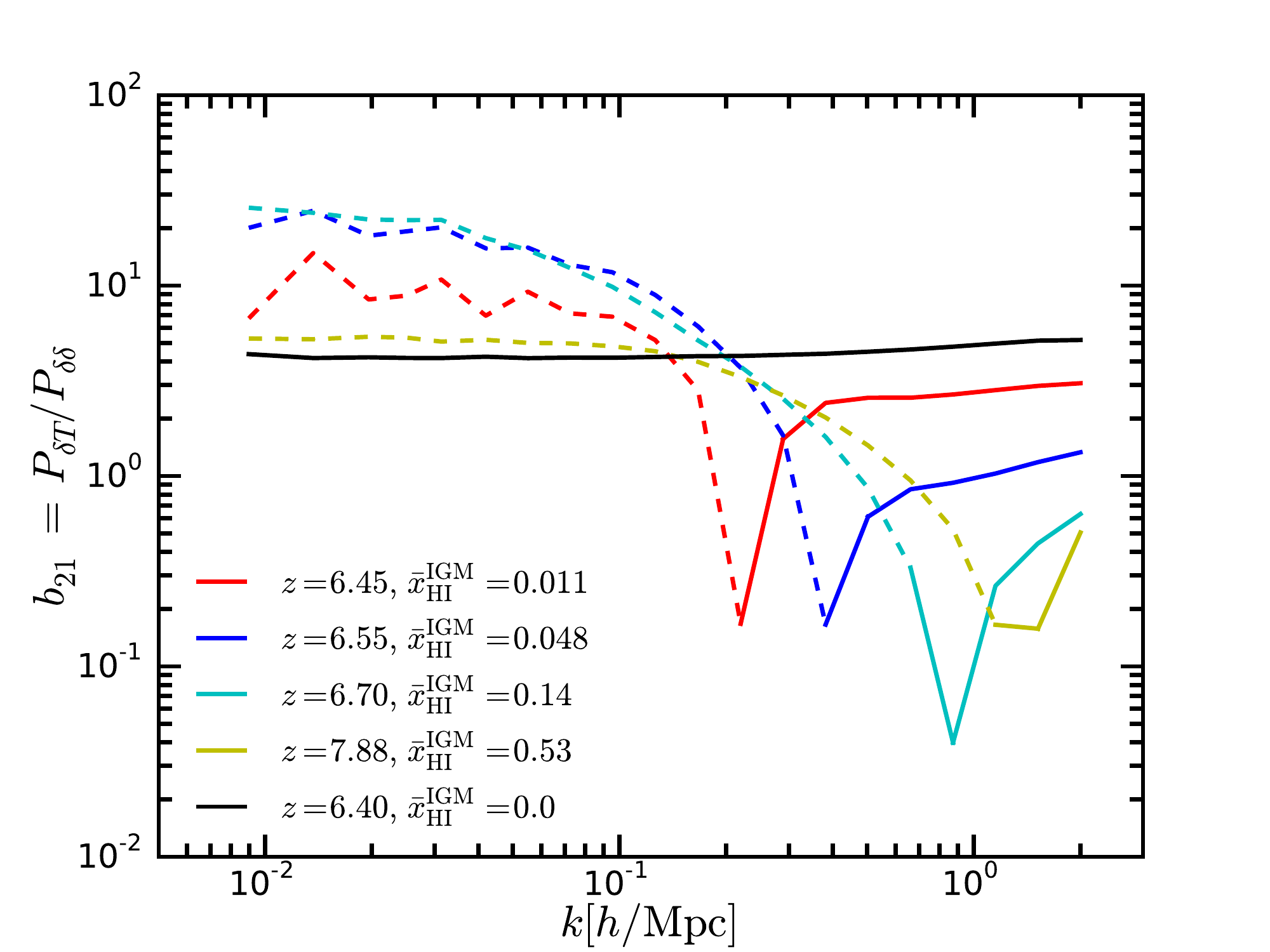}
\end{minipage}%
}%
\centering
\caption{The bias between the 21 cm brightness temperature and the dark matter density. 
The top and bottom panels show results during the early and late stage of reionization, respectively. 
}
\label{fig:bias2}
\end{figure}

\begin{figure}
\centering
\includegraphics[width=0.5\textwidth]{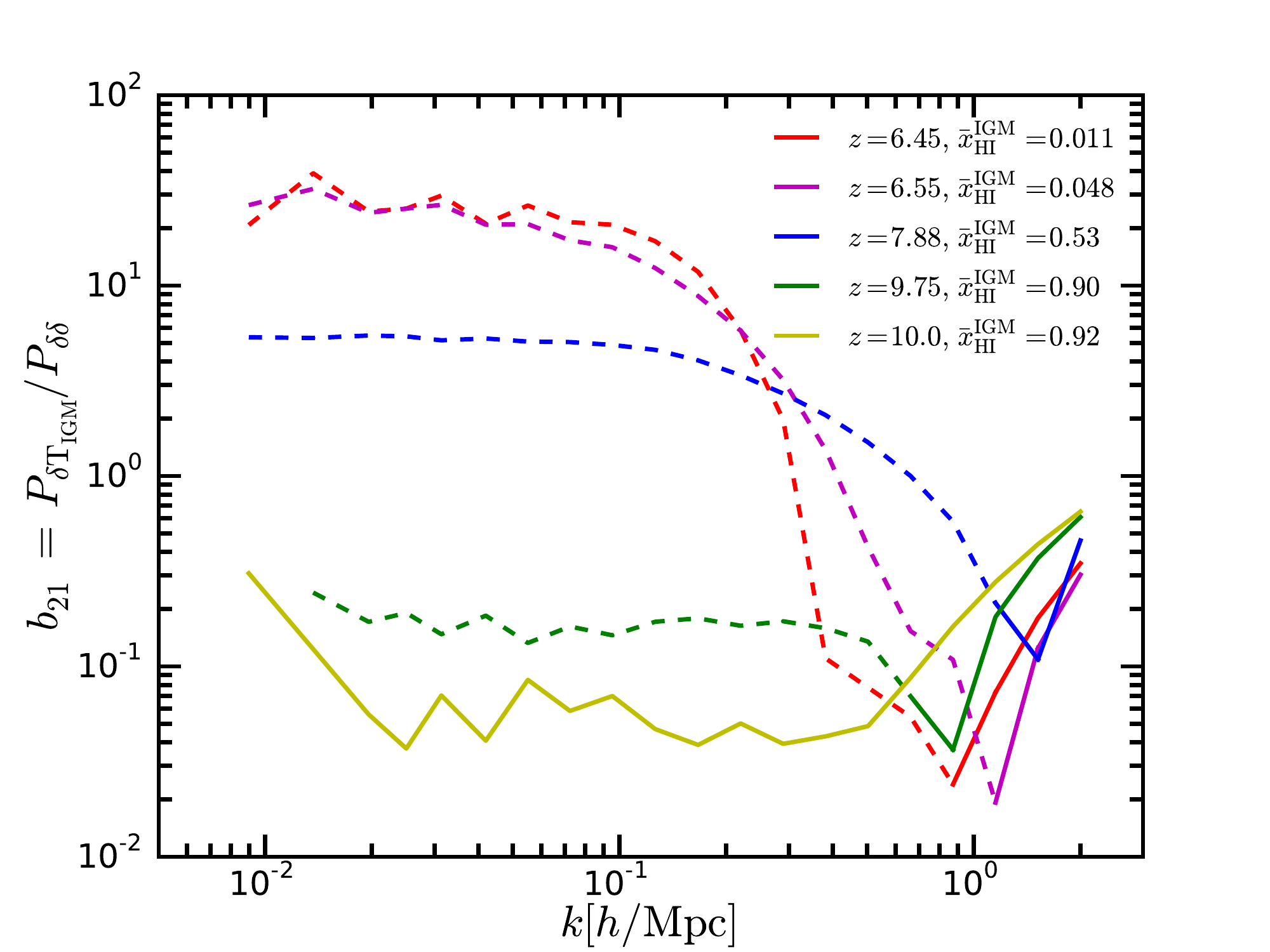}
\includegraphics[width=0.5\textwidth]{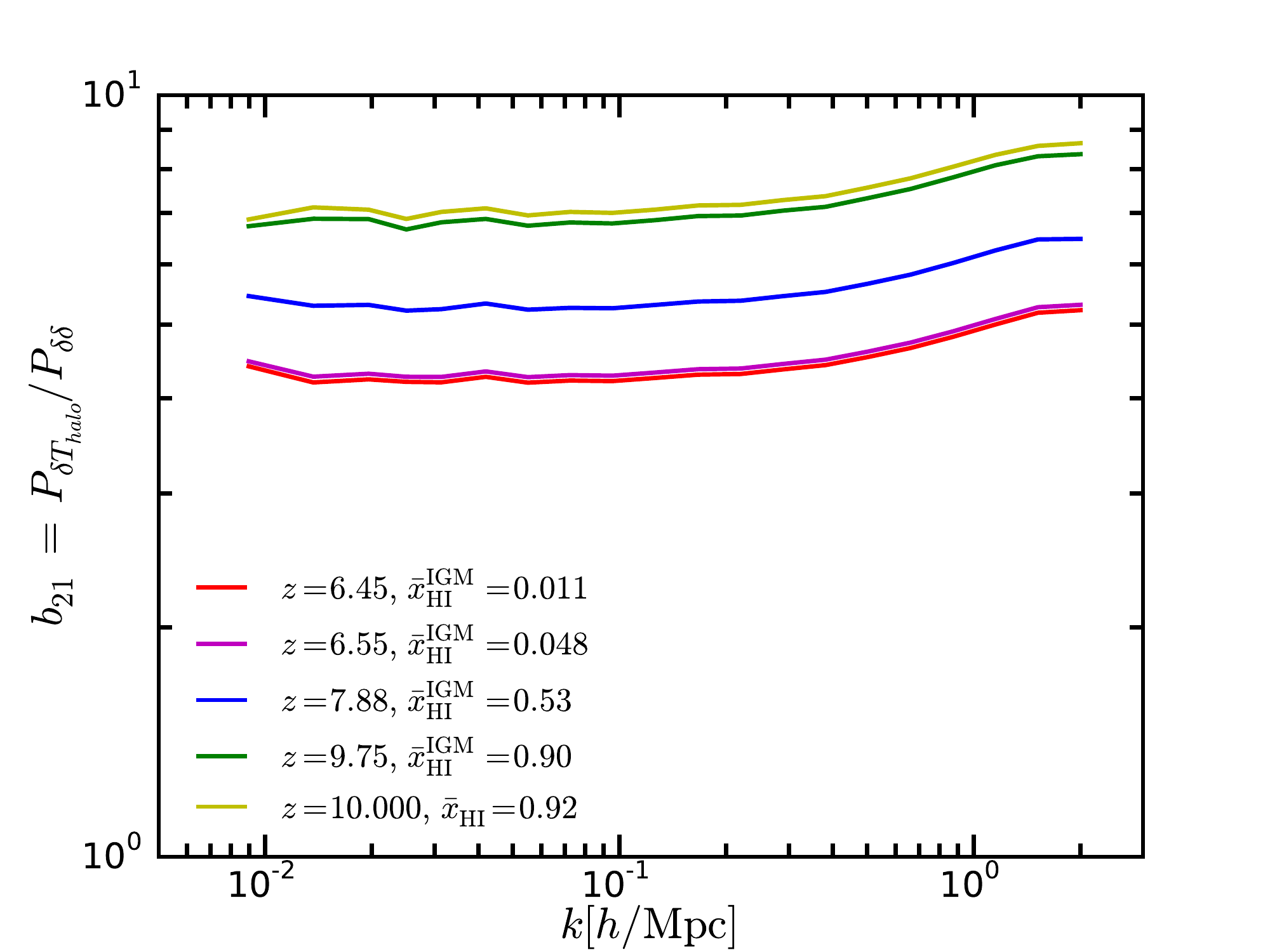}
\caption{The IGM (top panel) and halo (bottom panel) 21cm bias. }
\label{fig:deltahalobias}
\end{figure}

\subsection{The 21 cm bias $b_{21}$}

The evolution of the 21cm bias during the early and late stages of reionization is shown in Fig.~\ref{fig:bias2}. 
The error bars for the $z=9.90$ and $z=9.00$ curves are derived by similar method as Fig. \ref{fig:ps2}.
Unlike the neutral fraction bias, the 21cm bias shows more complicated behavior, 
especially in the early stage of EoR. The 21cm bias is always positive
on small scales, showing the dominance of the HI in halos. 
However the neutral fraction bias is totally negative in the small scales at early EoR, 
because the 21cm signal is weighted by local density, and the halo contribution is more apparent.

On large scales, the 21cm bias is positive at the beginning of reionization, then decreases to below zero and 
becomes more and more negative during the early EoR, and then
increases again and becomes less and less negative during the late EoR. 
During most of the EoR, there is a transition point at which the 21cm bias transits from 
negative at large scales to positive at small scales, and this transition scale only shows an obvious
increase in the late EoR. Perhaps due to this shift of transition scale,  the bias is scale-dependent up to fairly large scales during the late EoR.
We also show the $b_{\rm 21}$ after the completion of reionization, which is a positive
horizontal line in the figure, similar to the $b_f$.

We plot the 21cm bias for HI in halos as well as in the IGM in Fig.~\ref{fig:deltahalobias}. 
Similar to the cross-power spectra discussed in Sec. \ref{sec:ps}, on large scales the 21cm bias is dominated 
by the IGM while on small scales it is dominated by halos. A critical scale, below which halos contribute more to 
the 21cm bias than the IGM, increases with the decreasing redshift. Here we note that even considering only the 
IGM contribution, the 21cm bias is still positive on the smallest scales. This is due to the density fluctuations in the neutral regions: 
there are some relatively over-dense regions even in the under regions but not dense enough to collapse and produce ionizing photons. 

\begin{figure}
\centering\
\includegraphics[width=0.5\textwidth]{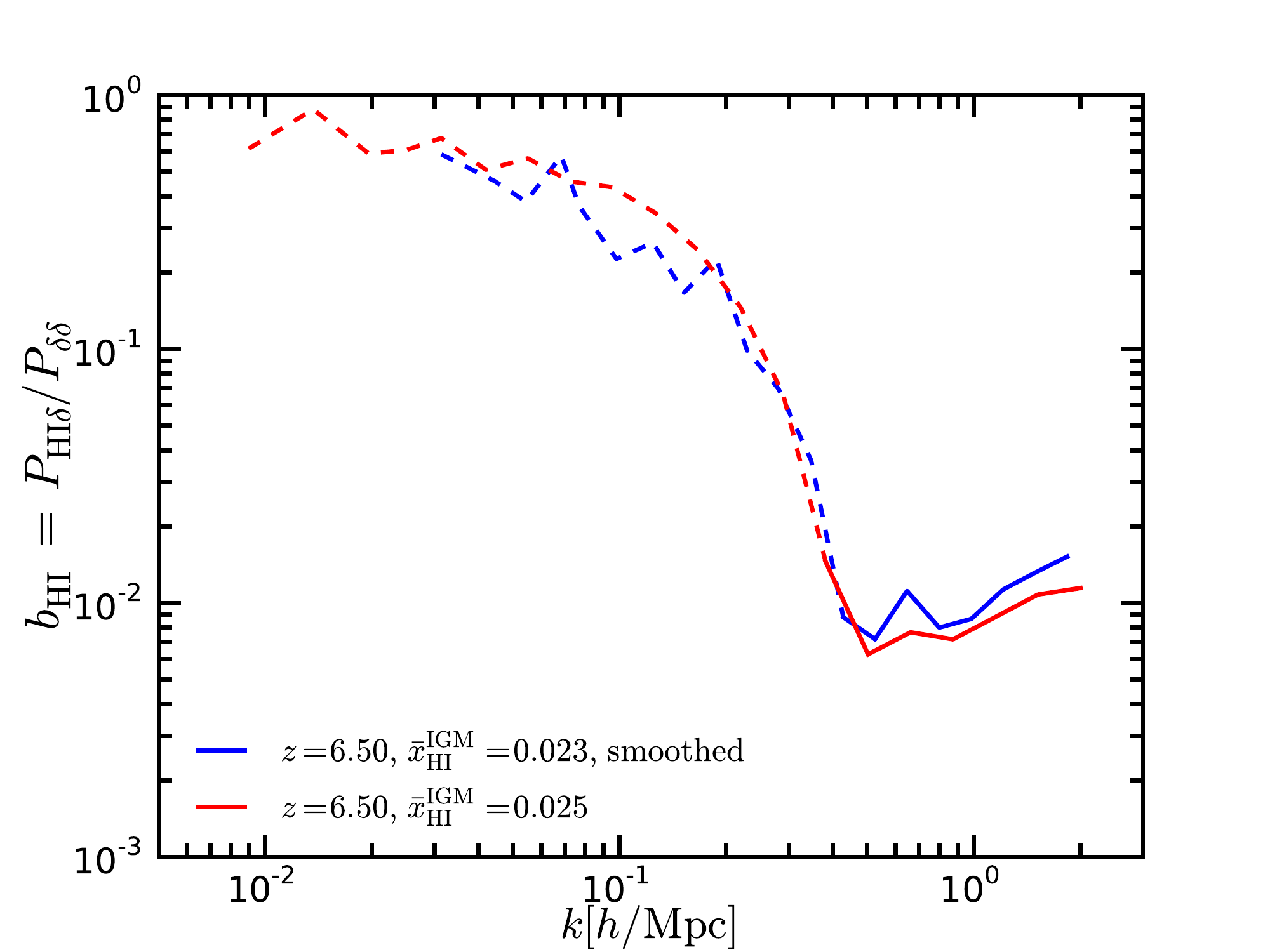}
\caption{The neutral fraction bias at $z = 6.50$. The red and blue lines show the result of the default
simulation and higher-resolution simulation ($512^3$ cells in a box of 286 Mpc side length) respectively. 
}
\label{fig:resolution1}
\end{figure}

We tested the resolution convergence of our simulation by using a simulation with higher resolution ($512^3$ cells) and 
smaller box size (286 Mpc per side) while keeping other parameters the same as our fiducial simulation. The
density and ionization fields  are smoothed to the same resolution as our fiducial simulation, then the 
halo HI content in each cell is assigned using the same model as described in section \ref{sec:massfunction}.
The neutral fraction bias of this higher resolution simulation and our fiducial simulation at the same redshift $z=6.50$ are shown 
in Fig.~\ref{fig:resolution1}. We see that they agree with each other. More importantly, the transition scales at which
the $b_f$ changes the sign agrees with each other very well, indicating the robustness of our conclusions.

\section{Conclusions and discussions}
\label{sec:conclusions}

In this paper, we investigated the HI distribution thoughout the EoR, including the HI in the IGM from semi-numerical simulations, and the HI in halos from an empirical relation between HI mass and its host halo mass.
We focused on the bias of the HI distribution with respect to the underlying matter density distribution,
in terms of the neutral fraction bias and the 21cm brightness temperature bias.

We found that during the reionization stage when $x_{\rm HI}^{\rm IGM}\gtrsim0.05$, the neutral fraction bias is always negative at all scales. However the amplitude first increases, up to the maximum $\sim 3$ when $x_{\rm HI}^{\rm IGM}\sim0.5$,  then decreases. When $x_{\rm HI}^{\rm IGM}\sim0.05$, positive bias appears at the smallest scales.
 Since then the negative-to-positive transition scale keeps increasing as the reionization approaches to the end in a short time scale. It is a good indicator of the completion of the reionization process.

The 21cm bias, however, is more complicated than the neutral fraction bias. Compared with neutral fraction, the 21cm brightness temperature is weighted by local gas density, therefore more correlated with the dark matter. At the smallest scales, the 21cm bias is always positive throughout the EoR, due to the contribution from HI in small neutral islands and in halos.
At large scales, the 21cm bias is positive at the early EoR stage,  however its 
amplitude decreases with decreasing IGM neutral fraction. At $x_{\rm HI}^{\rm IGM}\sim0.9$, the large scale 21cm bias transits from positive to negative. Since then the bias amplitude rises again, reaching its maximum absolute value 
when the IGM is roughly at $x_{\rm HI}^{\rm IGM}\sim0.1$.
From then on, the amplitude of the large scale 21cm bias drops again. However, at the same time the small scale 21cm bias always keeps increasing, because the contribution from HI in halos becomes more and more important. As a result, the transition point at which the 21cm bias transits from positive to negative, keeps increasing. Finally the reionization completed and all 21cm bias becomes positive again.

In summary,  a transition of 21cm bias from positive 
 to negative at large scales marks the starting of reionization, 
while 21cm bias reverts negative to positive on large scales with 
 the completion of reionization. 

The 21cm fluctuation could be measured with the current or upcoming generation experiments such as LOFAR, MWA,
HERA or SKA-low. The dark matter distribution and its cross-correlation with 21cm studied here can not be measured directly, 
but by cross-correlating the 21cm brightness field and some tracers of dark matter density, the 21cm bias can still be derived. The 
21cm-dark matter cross correlation and bias is however more general than specific tracers. A number of high redshift observations 
can be used to trace the large scale dark matter density,  for example the Lyman Break Galaxies (LBGs) and Ly$\alpha$ 
emitters (LAEs) derived from broadband or narrow band photometric observations
\citep{Dayal2012,Garel2015,Bouwens2015}. Of course, there will also some bias of the tracer sample with respect to the dark matter.
Indeed, \citet{Sobacchi2016} and \citet{Kubota2018} have investigated the detectability of 21cm-LAE cross-correlation
by combining the Subaru Hyper-Suprime Cam (HSC) Ultra-Deep field survey \citep{Aihara2018a,Aihara2018b} and 
the upcoming 21cm survey with LOFAR, MWA, or SKA-low. Measurement of the cross-correlation with the 
SKA-low is found very promising, and we expect the 21cm bias and its evolutionary features 
could be determined via such measurement in the SKA era.

\section*{Acknowledgements}
We thank Renyue Cen,  Kwan Chuen Chan  and Francisco Villaescusa-Navarro for helpful discussions. 
This work is supported by the the National Natural Science Foundation of China (NSFC) key project grant 11633004, 
the NSFC-ISF joint research program No. 11761141012, the CAS Frontier Science Key Project QYZDJ-SSW-SLH017, 
Chinese Academy of Sciences (CAS) Strategic Priority Research Program XDA15020200,   and
the MoST 2016YFE0100300. BY also acknowledges the support of  the CAS Pioneer Hundred Talents (Young Talents) program, 
the NSFC grant 11653003, the NSFC-CAS  joint fund for space scientific satellites No. U1738125.
ITI was supported by the Science and Technology Facilities Council [grant numbers ST/I000976/1, ST/F002858/1 
and ST/P000525/1];and The Southeast Physics Network (SEPNet). 
We acknowledge that the results in this paper have been achieved using the PRACE
Research Infrastructure resources Curie based at the Trs Grand Centre de Calcul (TGCC) operated by CEA, 
France and Marenostrum based in the Barcelona Supercomputing Center, Spain.The authors gratefully acknowledge 
the Gauss Centre for Supercomputing e.V.(www.gauss-centre.eu) for funding this project by providing computing time 
through the John von Neumann Institute for Computing
(NIC) on the GCS Supercomputers JURECA and JUWELS at Juelich Supercomputing Centre (JSC)



\bibliographystyle{mnras}
\bibliography{HI_bias}

\begin{thebibliography}{}
\makeatletter
\relax
\def\mn@urlcharsother{\let\do\@makeother \do\$\do\&\do\#\do\^\do\_\do\%\do\~}
\def\mn@doi{\begingroup\mn@urlcharsother \@ifnextchar [ {\mn@doi@}
  {\mn@doi@[]}}
\def\mn@doi@[#1]#2{\def\@tempa{#1}\ifx\@tempa\@empty \href
  {http://dx.doi.org/#2} {doi:#2}\else \href {http://dx.doi.org/#2} {#1}\fi
  \endgroup}
\def\mn@eprint#1#2{\mn@eprint@#1:#2::\@nil}
\def\mn@eprint@arXiv#1{\href {http://arxiv.org/abs/#1} {{\tt arXiv:#1}}}
\def\mn@eprint@dblp#1{\href {http://dblp.uni-trier.de/rec/bibtex/#1.xml}
  {dblp:#1}}
\def\mn@eprint@#1:#2:#3:#4\@nil{\def\@tempa {#1}\def\@tempb {#2}\def\@tempc
  {#3}\ifx \@tempc \@empty \let \@tempc \@tempb \let \@tempb \@tempa \fi \ifx
  \@tempb \@empty \def\@tempb {arXiv}\fi \@ifundefined
  {mn@eprint@\@tempb}{\@tempb:\@tempc}{\expandafter \expandafter \csname
  mn@eprint@\@tempb\endcsname \expandafter{\@tempc}}}

\bibitem[\protect\citeauthoryear{{Aihara} et~al.,}{{Aihara}
  et~al.}{2018a}]{Aihara2018a}
{Aihara} H.,  et~al., 2018a, \mn@doi [\pasj] {10.1093/pasj/psx066}, \href
  {http://adsabs.harvard.edu/abs/2018PASJ...70S...4A} {70, S4}

\bibitem[\protect\citeauthoryear{{Aihara} et~al.,}{{Aihara}
  et~al.}{2018b}]{Aihara2018b}
{Aihara} H.,  et~al., 2018b, \mn@doi [\pasj] {10.1093/pasj/psx081}, \href
  {http://adsabs.harvard.edu/abs/2018PASJ...70S...8A} {70, S8}

\bibitem[\protect\citeauthoryear{Alvarez et~al.}{Alvarez
  et~al.}{2019}]{Alvarez:2019pss}
Alvarez M.~A.,  et~al., 2019, arXiv e-prints (\mn@eprint {arXiv} {1903.04580})

\bibitem[\protect\citeauthoryear{{Bagla}, {Khandai}  \& {Datta}}{{Bagla}
  et~al.}{2010}]{Bagla2010}
{Bagla} J.~S.,  {Khandai} N.,   {Datta} K.~K.,  2010, \mn@doi [\mnras]
  {10.1111/j.1365-2966.2010.16933.x}, \href
  {http://adsabs.harvard.edu/abs/2010MNRAS.407..567B} {407, 567}

\bibitem[\protect\citeauthoryear{{Basilakos}, {Plionis}, {Kova{\v c}}  \&
  {Voglis}}{{Basilakos} et~al.}{2007}]{Basilakos2007}
{Basilakos} S.,  {Plionis} M.,  {Kova{\v c}} K.,   {Voglis} N.,  2007, \mn@doi
  [\mnras] {10.1111/j.1365-2966.2007.11781.x}, \href
  {http://adsabs.harvard.edu/abs/2007MNRAS.378..301B} {378, 301}

\bibitem[\protect\citeauthoryear{{Battaglia}, {Trac}, {Cen}  \&
  {Loeb}}{{Battaglia} et~al.}{2013}]{BTCL2013}
{Battaglia} N.,  {Trac} H.,  {Cen} R.,   {Loeb} A.,  2013, \mn@doi [\apj]
  {10.1088/0004-637X/776/2/81}, \href
  {http://adsabs.harvard.edu/abs/2013ApJ...776...81B} {776, 81}

\bibitem[\protect\citeauthoryear{{Bouwens} et~al.,}{{Bouwens}
  et~al.}{2015}]{Bouwens2015}
{Bouwens} R.~J.,  et~al., 2015, \mn@doi [\apj] {10.1088/0004-637X/803/1/34},
  \href {http://adsabs.harvard.edu/abs/2015ApJ...803...34B} {803, 34}

\bibitem[\protect\citeauthoryear{Bowman et~al.,}{Bowman
  et~al.}{2013}]{Bowman2013}
Bowman J.~D.,  et~al., 2013, \mn@doi [Publications of the Astronomical Society
  of Australia] {10.1017/pas.2013.009}, 30, e031

\bibitem[\protect\citeauthoryear{{Bull} et~al.,}{{Bull}
  et~al.}{2018}]{Bull2018}
{Bull} P.,  et~al., 2018, preprint, \href
  {https://ui.adsabs.harvard.edu/#abs/2018arXiv181002680B} {p.
  arXiv:1810.02680}

\bibitem[\protect\citeauthoryear{Chan, Hamaus  \& Desjacques}{Chan
  et~al.}{2014}]{Chan:2014qka}
Chan K.~C.,  Hamaus N.,   Desjacques V.,  2014, \mn@doi [Phys. Rev.]
  {10.1103/PhysRevD.90.103521}, D90, 103521

\bibitem[\protect\citeauthoryear{{Chen} \& {Miralda-Escud{\'e}}}{{Chen} \&
  {Miralda-Escud{\'e}}}{2004}]{2004ApJ...602....1C}
{Chen} X.,  {Miralda-Escud{\'e}} J.,  2004, \mn@doi [\apj] {10.1086/380829},
  \href {https://ui.adsabs.harvard.edu/abs/2004ApJ...602....1C} {602, 1}

\bibitem[\protect\citeauthoryear{{Chen} \& {Miralda-Escud{\'e}}}{{Chen} \&
  {Miralda-Escud{\'e}}}{2008}]{2008ApJ...684...18C}
{Chen} X.,  {Miralda-Escud{\'e}} J.,  2008, \mn@doi [\apj] {10.1086/528941},
  \href {https://ui.adsabs.harvard.edu/abs/2008ApJ...684...18C} {684, 18}

\bibitem[\protect\citeauthoryear{{Chen}, {Xu}, {Wang}  \& {Chen}}{{Chen}
  et~al.}{2018}]{Chen:2018enj}
{Chen} Z.,  {Xu} Y.,  {Wang} Y.,   {Chen} X.,  2018, arXiv e-prints, \href
  {https://ui.adsabs.harvard.edu/abs/2018arXiv181210333C} {p. arXiv:1812.10333}
  (\mn@eprint {arXiv} {1812.10333})

\bibitem[\protect\citeauthoryear{{Cooray} \& {Sheth}}{{Cooray} \&
  {Sheth}}{2002}]{Cooray2002}
{Cooray} A.,  {Sheth} R.,  2002, \mn@doi [\physrep]
  {10.1016/S0370-1573(02)00276-4}, \href
  {http://adsabs.harvard.edu/abs/2002PhR...372....1C} {372, 1}

\bibitem[\protect\citeauthoryear{{Dayal} \& {Ferrara}}{{Dayal} \&
  {Ferrara}}{2012}]{Dayal2012}
{Dayal} P.,  {Ferrara} A.,  2012, \mn@doi [\mnras]
  {10.1111/j.1365-2966.2012.20486.x}, \href
  {http://adsabs.harvard.edu/abs/2012MNRAS.421.2568D} {421, 2568}

\bibitem[\protect\citeauthoryear{DeBoer et~al.,}{DeBoer
  et~al.}{2017}]{DeBoer2017}
DeBoer D.~R.,  et~al., 2017, \mn@doi [Publications of the Astronomical Society
  of the Pacific] {10.1088/1538-3873/129/974/045001}, 129, 045001

\bibitem[\protect\citeauthoryear{{Dekel} \& {Lahav}}{{Dekel} \&
  {Lahav}}{1999}]{1999ApJ...520...24D}
{Dekel} A.,  {Lahav} O.,  1999, \mn@doi [\apj] {10.1086/307428}, \href
  {https://ui.adsabs.harvard.edu/abs/1999ApJ...520...24D} {520, 24}

\bibitem[\protect\citeauthoryear{{Dixon}, {Iliev}, {Mellema}, {Ahn}  \&
  {Shapiro}}{{Dixon} et~al.}{2016}]{2016MNRAS.456.3011D}
{Dixon} K.~L.,  {Iliev} I.~T.,  {Mellema} G.,  {Ahn} K.,   {Shapiro} P.~R.,
  2016, \mn@doi [\mnras] {10.1093/mnras/stv2887}, \href
  {https://ui.adsabs.harvard.edu/abs/2016MNRAS.456.3011D} {456, 3011}

\bibitem[\protect\citeauthoryear{{Emberson}, {Thomas}  \& {Alvarez}}{{Emberson}
  et~al.}{2013}]{Emberson2013}
{Emberson} J.~D.,  {Thomas} R.~M.,   {Alvarez} M.~A.,  2013, \mn@doi [\apj]
  {10.1088/0004-637X/763/2/146}, \href
  {http://adsabs.harvard.edu/abs/2013ApJ...763..146E} {763, 146}

\bibitem[\protect\citeauthoryear{{Furlanetto} \& {Oh}}{{Furlanetto} \&
  {Oh}}{2005}]{Furlanetto2005}
{Furlanetto} S.~R.,  {Oh} S.~P.,  2005, \mn@doi [\mnras]
  {10.1111/j.1365-2966.2005.09505.x}, \href
  {http://adsabs.harvard.edu/abs/2005MNRAS.363.1031F} {363, 1031}

\bibitem[\protect\citeauthoryear{{Furlanetto} \& {Oh}}{{Furlanetto} \&
  {Oh}}{2016}]{Furlanetto2016}
{Furlanetto} S.~R.,  {Oh} S.~P.,  2016, \mn@doi [\mnras]
  {10.1093/mnras/stw104}, \href
  {http://adsabs.harvard.edu/abs/2016MNRAS.457.1813F} {457, 1813}

\bibitem[\protect\citeauthoryear{{Furlanetto}, {Zaldarriaga}  \&
  {Hernquist}}{{Furlanetto} et~al.}{2004}]{Furlanetto2004}
{Furlanetto} S.~R.,  {Zaldarriaga} M.,   {Hernquist} L.,  2004, \mn@doi [\apj]
  {10.1086/423025}, \href {http://adsabs.harvard.edu/abs/2004ApJ...613....1F}
  {613, 1}

\bibitem[\protect\citeauthoryear{Furlanetto et~al.}{Furlanetto
  et~al.}{2019}]{Furlanetto:2019jzo}
Furlanetto S.,  et~al., 2019, arXiv e-prints (\mn@eprint {arXiv} {1903.06204})

\bibitem[\protect\citeauthoryear{{Garel}, {Blaizot}, {Guiderdoni},
  {Michel-Dansac}, {Hayes}  \& {Verhamme}}{{Garel} et~al.}{2015}]{Garel2015}
{Garel} T.,  {Blaizot} J.,  {Guiderdoni} B.,  {Michel-Dansac} L.,  {Hayes} M.,
   {Verhamme} A.,  2015, \mn@doi [\mnras] {10.1093/mnras/stv374}, \href
  {http://adsabs.harvard.edu/abs/2015MNRAS.450.1279G} {450, 1279}

\bibitem[\protect\citeauthoryear{{Giri}, {Mellema}, {Aldheimer}, {Dixon}  \&
  {Iliev}}{{Giri} et~al.}{2019}]{2019arXiv190301294G}
{Giri} S.~K.,  {Mellema} G.,  {Aldheimer} T.,  {Dixon} K.~L.,   {Iliev} I.~T.,
  2019, arXiv e-prints, \href
  {https://ui.adsabs.harvard.edu/abs/2019arXiv190301294G} {p. arXiv:1903.01294}
  (\mn@eprint {arXiv} {1903.01294})

\bibitem[\protect\citeauthoryear{{Gong}, {Chen}, {Silva}, {Cooray}  \&
  {Santos}}{{Gong} et~al.}{2011}]{Gong2011}
{Gong} Y.,  {Chen} X.,  {Silva} M.,  {Cooray} A.,   {Santos} M.~G.,  2011,
  \mn@doi [\apjl] {10.1088/2041-8205/740/1/L20}, \href
  {http://adsabs.harvard.edu/abs/2011ApJ...740L..20G} {740, L20}

\bibitem[\protect\citeauthoryear{{Guo}, {Li}, {Zheng}, {Mo}, {Jing}, {Zu},
  {Lim}  \& {Xu}}{{Guo} et~al.}{2017}]{Guo2017}
{Guo} H.,  {Li} C.,  {Zheng} Z.,  {Mo} H.~J.,  {Jing} Y.~P.,  {Zu} Y.,  {Lim}
  S.~H.,   {Xu} H.,  2017, \mn@doi [\apj] {10.3847/1538-4357/aa85e7}, \href
  {http://adsabs.harvard.edu/abs/2017ApJ...846...61G} {846, 61}

\bibitem[\protect\citeauthoryear{{Hoffmann}, {Mao}, {Xu}, {Mo}  \&
  {Wandelt}}{{Hoffmann} et~al.}{2018}]{Hoffmann2018}
{Hoffmann} K.,  {Mao} Y.,  {Xu} J.,  {Mo} H.,   {Wandelt} B.~D.,  2018,
  preprint, \href {https://ui.adsabs.harvard.edu/#abs/2018arXiv180202578H} {p.
  arXiv:1802.02578}

\bibitem[\protect\citeauthoryear{{Iliev}, {Mellema}, {Pen}, {Merz}, {Shapiro}
  \& {Alvarez}}{{Iliev} et~al.}{2006}]{Iliev2006}
{Iliev} I.~T.,  {Mellema} G.,  {Pen} U.-L.,  {Merz} H.,  {Shapiro} P.~R.,
  {Alvarez} M.~A.,  2006, \mn@doi [\mnras] {10.1111/j.1365-2966.2006.10502.x},
  \href {http://adsabs.harvard.edu/abs/2006MNRAS.369.1625I} {369, 1625}

\bibitem[\protect\citeauthoryear{{Iliev}, {Mellema}, {Ahn}, {Shapiro}, {Mao}
  \& {Pen}}{{Iliev} et~al.}{2014}]{2014MNRAS.439..725I}
{Iliev} I.~T.,  {Mellema} G.,  {Ahn} K.,  {Shapiro} P.~R.,  {Mao} Y.,   {Pen}
  U.-L.,  2014, \mn@doi [\mnras] {10.1093/mnras/stt2497}, \href
  {https://ui.adsabs.harvard.edu/abs/2014MNRAS.439..725I} {439, 725}

\bibitem[\protect\citeauthoryear{{Kim}, {Wyithe}, {Park}, {Poole}, {Lacey}  \&
  {Baugh}}{{Kim} et~al.}{2016}]{2016MNRAS.455.4498K}
{Kim} H.-S.,  {Wyithe} J. S.~B.,  {Park} J.,  {Poole} G.~B.,  {Lacey} C.~G.,
  {Baugh} C.~M.,  2016, \mn@doi [\mnras] {10.1093/mnras/stv2623}, \href
  {https://ui.adsabs.harvard.edu/abs/2016MNRAS.455.4498K} {455, 4498}

\bibitem[\protect\citeauthoryear{{Koopmans} et~al.,}{{Koopmans}
  et~al.}{2015}]{Koopmans2015}
{Koopmans} L.,  et~al., 2015, Advancing Astrophysics with the Square Kilometre
  Array (AASKA14), \href
  {https://ui.adsabs.harvard.edu/abs/2015aska.confE...1K} {p.~1}

\bibitem[\protect\citeauthoryear{{Kubota}, {Yoshiura}, {Takahashi}, {Hasegawa},
  {Yajima}, {Ouchi}, {Pindor}  \& {Webster}}{{Kubota}
  et~al.}{2018}]{Kubota2018}
{Kubota} K.,  {Yoshiura} S.,  {Takahashi} K.,  {Hasegawa} K.,  {Yajima} H.,
  {Ouchi} M.,  {Pindor} B.,   {Webster} R.~L.,  2018, \mn@doi [\mnras]
  {10.1093/mnras/sty1471}, \href
  {http://adsabs.harvard.edu/abs/2018MNRAS.479.2754K} {479, 2754}

\bibitem[\protect\citeauthoryear{Liu et~al.}{Liu et~al.}{2019}]{Liu:2019srd}
Liu A.,  et~al., 2019, arXiv e-prints (\mn@eprint {arXiv} {1903.06240})

\bibitem[\protect\citeauthoryear{{Mar{\'{\i}}n}, {Gnedin}, {Seo}  \&
  {Vallinotto}}{{Mar{\'{\i}}n} et~al.}{2010}]{Marn2010}
{Mar{\'{\i}}n} F.~A.,  {Gnedin} N.~Y.,  {Seo} H.-J.,   {Vallinotto} A.,  2010,
  \mn@doi [\apj] {10.1088/0004-637X/718/2/972}, \href
  {http://adsabs.harvard.edu/abs/2010ApJ...718..972M} {718, 972}

\bibitem[\protect\citeauthoryear{{Martin}, {Giovanelli}, {Haynes}  \&
  {Guzzo}}{{Martin} et~al.}{2012}]{Martin2012}
{Martin} A.~M.,  {Giovanelli} R.,  {Haynes} M.~P.,   {Guzzo} L.,  2012, \mn@doi
  [\apj] {10.1088/0004-637X/750/1/38}, \href
  {http://adsabs.harvard.edu/abs/2012ApJ...750...38M} {750, 38}

\bibitem[\protect\citeauthoryear{{McQuinn} \& {D'Aloisio}}{{McQuinn} \&
  {D'Aloisio}}{2018}]{McQuinn2018}
{McQuinn} M.,  {D'Aloisio} A.,  2018, \mn@doi [Journal of Cosmology and
  Astro-Particle Physics] {10.1088/1475-7516/2018/10/016}, \href
  {https://ui.adsabs.harvard.edu/abs/2018JCAP...10..016M} {2018, 016}

\bibitem[\protect\citeauthoryear{{McQuinn}, {Oh}  \&
  {Faucher-Gigu{\`e}re}}{{McQuinn} et~al.}{2011}]{McQuinn2011}
{McQuinn} M.,  {Oh} S.~P.,   {Faucher-Gigu{\`e}re} C.-A.,  2011, \mn@doi [\apj]
  {10.1088/0004-637X/743/1/82}, \href
  {http://adsabs.harvard.edu/abs/2011ApJ...743...82M} {743, 82}

\bibitem[\protect\citeauthoryear{{Mesinger} \& {Furlanetto}}{{Mesinger} \&
  {Furlanetto}}{2007}]{Mesinger2007}
{Mesinger} A.,  {Furlanetto} S.,  2007, \mn@doi [\apj] {10.1086/521806}, \href
  {http://adsabs.harvard.edu/abs/2007ApJ...669..663M} {669, 663}

\bibitem[\protect\citeauthoryear{{Mesinger}, {Furlanetto}  \& {Cen}}{{Mesinger}
  et~al.}{2011}]{Mesinger2011}
{Mesinger} A.,  {Furlanetto} S.,   {Cen} R.,  2011, \mn@doi [\mnras]
  {10.1111/j.1365-2966.2010.17731.x}, \href
  {http://adsabs.harvard.edu/abs/2011MNRAS.411..955M} {411, 955}

\bibitem[\protect\citeauthoryear{{Obuljen}, {Alonso}, {Villaescusa-Navarro},
  {Yoon}  \& {Jones}}{{Obuljen} et~al.}{2018a}]{Obuljen2018a}
{Obuljen} A.,  {Alonso} D.,  {Villaescusa-Navarro} F.,  {Yoon} I.,   {Jones}
  M.,  2018a, arXiv e-prints, \href
  {http://adsabs.harvard.edu/abs/2018arXiv180500934O} {} (\mn@eprint {arXiv}
  {1805.00934})

\bibitem[\protect\citeauthoryear{{Obuljen}, {Castorina}, {Villaescusa-Navarro}
  \& {Viel}}{{Obuljen} et~al.}{2018b}]{Obuljen2018b}
{Obuljen} A.,  {Castorina} E.,  {Villaescusa-Navarro} F.,   {Viel} M.,  2018b,
  \mn@doi [\jcap] {10.1088/1475-7516/2018/05/004}, \href
  {http://adsabs.harvard.edu/abs/2018JCAP...05..004O} {5, 004}

\bibitem[\protect\citeauthoryear{{Padmanabhan} \& {Refregier}}{{Padmanabhan} \&
  {Refregier}}{2017}]{Padmanabhan2017}
{Padmanabhan} H.,  {Refregier} A.,  2017, \mn@doi [\mnras]
  {10.1093/mnras/stw2706}, \href
  {http://adsabs.harvard.edu/abs/2017MNRAS.464.4008P} {464, 4008}

\bibitem[\protect\citeauthoryear{{Parsons} et~al.,}{{Parsons}
  et~al.}{2010}]{Parsons2010}
{Parsons} A.~R.,  et~al., 2010, \mn@doi [\aj] {10.1088/0004-6256/139/4/1468},
  \href {https://ui.adsabs.harvard.edu/abs/2010AJ....139.1468P} {139, 1468}

\bibitem[\protect\citeauthoryear{{Peterson} et~al.,}{{Peterson}
  et~al.}{2009}]{Peterson2009}
{Peterson} J.~B.,  et~al., 2009, in astro2010: The Astronomy and Astrophysics
  Decadal Survey.  (\mn@eprint {arXiv} {0902.3091})

\bibitem[\protect\citeauthoryear{{Popping}, {Behroozi}  \& {Peeples}}{{Popping}
  et~al.}{2015}]{Popping2015}
{Popping} G.,  {Behroozi} P.~S.,   {Peeples} M.~S.,  2015, \mn@doi [\mnras]
  {10.1093/mnras/stv318}, \href
  {http://adsabs.harvard.edu/abs/2015MNRAS.449..477P} {449, 477}

\bibitem[\protect\citeauthoryear{{Pritchard} \& {Loeb}}{{Pritchard} \&
  {Loeb}}{2012}]{Pritchard2012}
{Pritchard} J.~R.,  {Loeb} A.,  2012, \mn@doi [Reports on Progress in Physics]
  {10.1088/0034-4885/75/8/086901}, \href
  {http://adsabs.harvard.edu/abs/2012RPPh...75h6901P} {75, 086901}

\bibitem[\protect\citeauthoryear{{Santos}, {Ferramacho}, {Silva}, {Amblard}  \&
  {Cooray}}{{Santos} et~al.}{2010}]{2010MNRAS.406.2421S}
{Santos} M.~G.,  {Ferramacho} L.,  {Silva} M.~B.,  {Amblard} A.,   {Cooray} A.,
   2010, \mn@doi [\mnras] {10.1111/j.1365-2966.2010.16898.x}, \href
  {https://ui.adsabs.harvard.edu/abs/2010MNRAS.406.2421S} {406, 2421}

\bibitem[\protect\citeauthoryear{{Sarkar}, {Bharadwaj}  \&
  {Anathpindika}}{{Sarkar} et~al.}{2016}]{Sarkar2016}
{Sarkar} D.,  {Bharadwaj} S.,   {Anathpindika} S.,  2016, \mn@doi [\mnras]
  {10.1093/mnras/stw1111}, \href
  {http://adsabs.harvard.edu/abs/2016MNRAS.460.4310S} {460, 4310}

\bibitem[\protect\citeauthoryear{{Sheth} \& {Tormen}}{{Sheth} \&
  {Tormen}}{1999}]{Sheth1999}
{Sheth} R.~K.,  {Tormen} G.,  1999, \mn@doi [\mnras]
  {10.1046/j.1365-8711.1999.02692.x}, \href
  {http://adsabs.harvard.edu/abs/1999MNRAS.308..119S} {308, 119}

\bibitem[\protect\citeauthoryear{Sheth \& van~de Weygaert}{Sheth \& van~de
  Weygaert}{2004}]{Sheth:2003py}
Sheth R.~K.,  van~de Weygaert R.,  2004, \mn@doi [Mon. Not. Roy. Astron. Soc.]
  {10.1111/j.1365-2966.2004.07661.x}, 350, 517

\bibitem[\protect\citeauthoryear{{Sobacchi}, {Mesinger}  \& {Greig}}{{Sobacchi}
  et~al.}{2016}]{Sobacchi2016}
{Sobacchi} E.,  {Mesinger} A.,   {Greig} B.,  2016, \mn@doi [\mnras]
  {10.1093/mnras/stw811}, \href
  {http://adsabs.harvard.edu/abs/2016MNRAS.459.2741S} {459, 2741}

\bibitem[\protect\citeauthoryear{{Songaila} \& {Cowie}}{{Songaila} \&
  {Cowie}}{2010}]{2010ApJ...721.1448S}
{Songaila} A.,  {Cowie} L.~L.,  2010, \mn@doi [\apj]
  {10.1088/0004-637X/721/2/1448}, \href
  {http://adsabs.harvard.edu/abs/2010ApJ...721.1448S} {721, 1448}

\bibitem[\protect\citeauthoryear{Tingay et~al.,}{Tingay
  et~al.}{2013}]{Tingay2013}
Tingay S.~J.,  et~al., 2013, \mn@doi [Publications of the Astronomical Society
  of Australia] {10.1017/pasa.2012.007}, 30, e007

\bibitem[\protect\citeauthoryear{{Trac} \& {Cen}}{{Trac} \&
  {Cen}}{2007}]{Trac2007}
{Trac} H.,  {Cen} R.,  2007, \mn@doi [\apj] {10.1086/522566}, \href
  {http://adsabs.harvard.edu/abs/2007ApJ...671....1T} {671, 1}

\bibitem[\protect\citeauthoryear{{Villaescusa-Navarro}
  et~al.,}{{Villaescusa-Navarro} et~al.}{2018}]{Navarro2018}
{Villaescusa-Navarro} F.,  et~al., 2018, \mn@doi [\apj]
  {10.3847/1538-4357/aadba0}, \href
  {https://ui.adsabs.harvard.edu/abs/2018ApJ...866..135V} {866, 135}

\bibitem[\protect\citeauthoryear{{Wang} et~al.,}{{Wang}
  et~al.}{2019}]{Wang2019}
{Wang} Z.,  et~al., 2019, arXiv e-prints, \href
  {https://ui.adsabs.harvard.edu/abs/2019arXiv190102724W} {p. arXiv:1901.02724}
  (\mn@eprint {arXiv} {1901.02724})

\bibitem[\protect\citeauthoryear{{Xu}, {Yue}, {Su}, {Fan}  \& {Chen}}{{Xu}
  et~al.}{2014}]{Xu2014}
{Xu} Y.,  {Yue} B.,  {Su} M.,  {Fan} Z.,   {Chen} X.,  2014, \mn@doi [\apj]
  {10.1088/0004-637X/781/2/97}, \href
  {http://adsabs.harvard.edu/abs/2014ApJ...781...97X} {781, 97}

\bibitem[\protect\citeauthoryear{{Xu}, {Wang}  \& {Chen}}{{Xu}
  et~al.}{2015}]{Xu2015A}
{Xu} Y.,  {Wang} X.,   {Chen} X.,  2015, \mn@doi [\apj]
  {10.1088/0004-637X/798/1/40}, \href
  {https://ui.adsabs.harvard.edu/#abs/2015ApJ...798...40X} {798, 40}

\bibitem[\protect\citeauthoryear{{Xu}, {Hamann}  \& {Chen}}{{Xu}
  et~al.}{2016}]{Xu2016A}
{Xu} Y.,  {Hamann} J.,   {Chen} X.,  2016, \mn@doi [\prd]
  {10.1103/PhysRevD.94.123518}, \href
  {https://ui.adsabs.harvard.edu/#abs/2016PhRvD..94l3518X} {94, 123518}

\bibitem[\protect\citeauthoryear{{Xu}, {Yue}  \& {Chen}}{{Xu}
  et~al.}{2017}]{Xu2017}
{Xu} Y.,  {Yue} B.,   {Chen} X.,  2017, \mn@doi [\apj]
  {10.3847/1538-4357/aa7b71}, \href
  {http://adsabs.harvard.edu/abs/2017ApJ...844..117X} {844, 117}

\bibitem[\protect\citeauthoryear{{Yue}, {Ciardi}, {Scannapieco}  \&
  {Chen}}{{Yue} et~al.}{2009}]{Yue2009}
{Yue} B.,  {Ciardi} B.,  {Scannapieco} E.,   {Chen} X.,  2009, \mn@doi [\mnras]
  {10.1111/j.1365-2966.2009.15261.x}, \href
  {http://adsabs.harvard.edu/abs/2009MNRAS.398.2122Y} {398, 2122}

\bibitem[\protect\citeauthoryear{{Zahn}, {Lidz}, {McQuinn}, {Dutta},
  {Hernquist}, {Zaldarriaga}  \& {Furlanetto}}{{Zahn} et~al.}{2007}]{Zahn2007}
{Zahn} O.,  {Lidz} A.,  {McQuinn} M.,  {Dutta} S.,  {Hernquist} L.,
  {Zaldarriaga} M.,   {Furlanetto} S.~R.,  2007, \mn@doi [\apj]
  {10.1086/509597}, \href {http://adsabs.harvard.edu/abs/2007ApJ...654...12Z}
  {654, 12}

\bibitem[\protect\citeauthoryear{{Zahn}, {Mesinger}, {McQuinn}, {Trac}, {Cen}
  \& {Hernquist}}{{Zahn} et~al.}{2011}]{Zahn2011}
{Zahn} O.,  {Mesinger} A.,  {McQuinn} M.,  {Trac} H.,  {Cen} R.,   {Hernquist}
  L.~E.,  2011, \mn@doi [\mnras] {10.1111/j.1365-2966.2011.18439.x}, \href
  {http://adsabs.harvard.edu/abs/2011MNRAS.414..727Z} {414, 727}

\bibitem[\protect\citeauthoryear{{van Haarlem} et~al.,}{{van Haarlem}
  et~al.}{2013}]{vanHaarlem2013}
{van Haarlem} M.~P.,  et~al., 2013, \mn@doi [\aap]
  {10.1051/0004-6361/201220873}, \href
  {https://ui.adsabs.harvard.edu/abs/2013A%26A...556A...2V} {556, A2}

\makeatother
\end{thebibliography}

\label{lastpage}
\end{document}